\documentclass[10pt,aps,prx,twocolumn,english,superscriptaddress,floatfix,longbibliography]{revtex4-2}

\usepackage{enumitem}
\usepackage{graphicx}% Include figure files
\usepackage{dcolumn}% Align table columns on decimal point
\usepackage{bm}% bold math

\usepackage{amssymb}
\usepackage{lmodern}
\usepackage[utf8]{inputenc}
\usepackage{color}

\usepackage{varioref}
\usepackage{float}
\usepackage{units}
\usepackage{mathtools}
\usepackage{enumitem}

 \usepackage{amsmath}
\usepackage{amsthm}
\usepackage{amssymb}
\usepackage{mathdots}
\usepackage{graphicx}

\usepackage{array}
\usepackage{calc}
\usepackage[table]{xcolor}

\usepackage{multirow} %allows multiple rows in tables tabular

\usepackage{bbm} %allows for fancy 1, i.e. mathbb{1}, using \mathbbm{1}

\usepackage[unicode=true,pdfusetitle,
 bookmarks=true,bookmarksnumbered=false,bookmarksopen=false,
 breaklinks=true,pdfborder={0 0 0},pdfborderstyle={},backref=false,colorlinks=true]{hyperref}
\newcommand{\ket}  [1]{\left|#1\right\rangle}
\newcommand{\bra}  [1]{\left\langle#1\right|}

\newlength\celldim
\newlength\fontheight
\newlength\extraheight
\newcolumntype{S}
{ @{}
>{\centering\arraybackslash}
p{\celldim}
<{\rule[-0.2\extraheight]{0pt}%
{\fontheight + \extraheight/4}}
@{} }

\begin{document}
\preprint{APS/123-QED}
\title{Topological Tenfold Classification and the Entanglement of Two Qubits
%Approaching Two Qubits Entanglement using Topological Tenfold Classification 
%Quantum Entanglement in Two Qubits: topological properties and tenfold classification
%Topological Aspects of Quantum Entanglement in Two Qubit Systems
%Ver1: Topology and Entanglement in \\Two-Qubit Analogous Quantum Condensed Matter
%Ver0: Concurrence, Topology and Concurrence
}
\author{Nadav Orion}
\affiliation{%
Department of Physics, Technion -- Israel Institute of Technology, Haifa 3200003, Israel
}%
\author{Eric Akkermans}%
% \email{eric@physics.technion.ac.il}
\affiliation{%
Department of Physics, Technion -- Israel Institute of Technology, Haifa 3200003, Israel
}%
\date{\today}% It is always \today, today,
\begin{abstract}
%Topological phases of insulators and superconductors constitute an interesting platform to realize entangled quantum states. 
We present a constructive method utilizing the Cartan decomposition to characterize topological properties and their connection to two-qubit quantum entanglement, in the framework of the tenfold classification and Wootters' concurrence. This relationship is comprehensively established for the 2-qubit system through the antiunitary time reversal (TR) operator. The TR operator is shown to identify concurrence and differentiate between entangling and non-entangling operators. This distinction is of a topological nature, as the inclusion or exclusion of certain operators alters  topological characteristics. Proofs are presented which demonstrate that the 2-qubit system can be described in the framework of the tenfold classification, unveiling aspects of the  connection between entanglement and a geometrical phase. Topological features are obtained systematically by a mapping to a quantum graph, allowing for a direct computation of topological integers and of the 2-qubit equivalent of topological zero-modes. An additional perspective is provided regarding the extension of this new approach to condensed matter systems, illustrated through examples involving indistinguishable fermions and arrays of quantum dots.

%We offer a constructive approach based on the Cartan decomposition to describe the topological features and their relation to quantum entanglement. The Cartan decomposition classifies properties of Lie algebras by antiunitary operators. In condensed matter this is known as the tenfold way, which utilizes antiunitary symmetries to determine topological invariants. In entanglement theory this is known as the Wootters concurrence and 

\end{abstract}
\maketitle
%
%\tableofcontents

%%%%%%%%%%%%%%%%%%%%%%%%%%%%%%%%%%%%%%%%%%%%%%%%%
%%%%%%%%%%%%%%%%%%%%%%%%%%%%%%%%%%%%%%%%%%%%%%%%%

\section{Introduction}
The search for quantum entanglement \cite{Haroche2006,Horodecki2009,Paneru2020} between distinguishable qubits and its implementation in measurable physical setups constitutes an active field in physics and technology. Condensed matter systems are serious contenders to realize quantum entanglement since they provide ideal platforms to create qubits and to study their correlations \cite{Chtchelkatchev2002,Beenakker2003,Costa2006,Beenakker2006,Vandersypen2007,Schomerus2007,Hannes2008,doucot2008,Samuelsson2009,Rohling2012,Roche2020}. However, these systems are notoriously difficult to control. For instance, their intrinsic many-body nature makes it challenging to identify relevant degrees of freedom, either theoretically or experimentally. A possible way out is to harness isolated effective excitations such as topologically protected states.

The role of topology is essential in the contemporary understanding of band insulators \cite{TKNN,Altland1997,Kitaev2009,Stone2010,Ludwig2016,Chiu2016,Pires_Topology_book}. Topological invariants, like Chern numbers, indicate the presence of edge states, thus offering intriguing avenues for the creation of topologically protected qubits that remain stable against disturbances such as disorder. %\cite{Akkermans2007}.

This paper aims to demonstrate that the concepts of entanglement and the aforementioned topological considerations are not just convenient but inherently linked. Unlike earlier studies on this connection \cite{Kus2001,Avron2009,Chen2010}, we offer a different perspective, utilizing antiunitary symmetries and Cartan decomposition. The Cartan classification of symmetric spaces \cite{Helgason1978} is a powerful tool used to identify topological features of specific spaces, or the lack thereof. In condensed matter, the Cartan classification underlies the tenfold classification of insulators and superconductors \cite{Altland1997,Kitaev2009,Stone2010,Ludwig2016}, which determines topological characteristics of weakly interacting fermions. The identification of the (Cartan) symmetry class and topological characteristics of the relevant system, as a function of  spatial dimension, is achieved by employing two antiunitary symmetries: time reversal ($\Theta$) and particle-hole ($\Pi$).

In this paper, we consider physical systems that amount to two distinguishable qubits, "Alice and Bob". To characterize their entanglement in a state $\ket{\psi} $, we define the Wootters concurrence \cite{Hill1997,Wootters1998} by
%We wish to measure and change at will entanglement properties of two distinguishable qubits. The purpose of this paper is to fulfill this program with the help of topological properties. More specifically, entanglement of two qubits in a state $\ket{\psi} $ is characterized by its Wootters concurrence \cite{Hill1997,Wootters1998}, defined by 
\begin{equation}
    C_\psi \equiv\left|\bra{\psi}\Theta\ket{\psi}\right|, \label{eq:concurrence}
\end{equation} 
where $\Theta$ is an antiunitary symmetry of the two-qubit system. The Wootters concurrence fully characterizes quantum entanglement, and it also unveils topological features with the help of the Cartan decomposition: Alice and Bob go topological.

The intrinsic finiteness of 2-qubit systems extends topology beyond the tenfold classification used for condensed matter systems, which generally presuppose the thermodynamic limit. This assumption significantly narrows the variety of possible topological phases. However, in the case of two-qubit systems, such a simplification is inapplicable. We argue that the existence and efficiency of quantum entanglement are linked to a non-trivial topology. Stated otherwise, 2-qubit systems that lack topological features cannot be entangled. 
%We thus argue entangling operators are the result of non-trivial topology. Hence, no topological phenomena means there can be no consistent entanglement creation. The most natural systems to describe entanglement are therefore those where topology compels Hamiltonians to entangle.  

Equipped with a measure of entanglement, we wish to examine the conditions under which a time evolution, i.e. quantum gates, increases the concurrence of a state $\ket{\psi (t)} = U(t) \ket{\psi (0)}$, using a Hamiltonian $H$ and $U(t) = e^{-iHt}$. A challenging problem is to optimize the quantum gates, hence $U(t)$, to maximize concurrence and minimize the time required to achieve it. These questions have been addressed using a wealth of methods. Here, we wish to show that topological features of the Hilbert space of two %distinguishable 
qubits are an asset to achieve that purpose in a useful and original way.

The time evolution of a two-qubit system is described by an operator $U(t)$ in the $U(4)$ Lie group. It turns out that operators local in the product Hilbert space of two qubits leave entanglement unchanged. The classification of all non-local hence entangling operators has been achieved, and it relies on the Cartan decomposition of Hamiltonians $H$, hence of the Lie algebra $\mathfrak{su}(4)$ associated to the Lie group $U(4)$. This decomposition is presented in Section \ref{sec:Cartan}.

The classification of Hamiltonians and their Lie algebra using the antiunitary symmetry $\Theta$ is a part of a more general classification of symmetric Lie algebras, a.k.a. tenfold classification. The tenfold classification is based on two antiunitary symmetries, time-reversal $(\Theta)$ and particle-hole $(\Pi)$, and it uses them to identify stable spaces for Hamiltonians, unitary evolution operators $U(t)$, and their topology, e.g. their connectedness. For finite systems, these spaces are no longer stable, so topology is more involved. This is discussed in Section \ref{sec:Topology}.

By applying these topological considerations to two-qubit models, we establish straightforward connections to our established concepts of concurrence and entangling operators, as elaborated in Section \ref{sec:2qubits}.

Section \ref{sec:topo_phases} explores the shared Cartan classification between two qubits and condensed matter systems. We reformulate the non-local entangling Hamiltonians using a tight-binding approach to methodically identify topological phases and examine potential phase transitions.

Section \ref{sec:generalizations} explores how these findings can be extended to quantum computing scenarios involving more than two qubits, including indistinguishable fermions within condensed matter systems.

%This makes generalizing the results presented here more straightforward, and simple examples are noted the section \ref{sec:generalizations}.

%%%%%%%%%%%%%%%%%%%%%%%%%%%%%%%%%%%%%%%%%%%%%%%%%%%%%%%%%%%%%%%%%%%%%%%%%%%%%
%%%%%%%%%%%%%%%%%%%%%%%%%%%%%%%%%%%%%%%%%%%%%%%%%%%%%%%%%%%%%%%%%%%
%%%%%%%%%%%%%%%%%%%%%%%%%%%%%%%%%%%%%%%%%%%%%%%%%%%%%%%%%%%%%%%%%%%

\section{Cartan Decomposition and Locality \label{sec:Cartan}}

The Cartan decomposition \cite{Helgason1978} divides a real Lie algebra $\mathfrak{g}$ into two components, $\mathfrak{g}=\mathfrak{h}\oplus \mathfrak{p}$, distinguished by the following commutation relations:
\begin{equation}
 [h,h]\in \mathfrak{h},\quad [p,p]\in \mathfrak{h},\quad [h,p]\in \mathfrak{p}   .
 \label{eq:Cartan_Decomp}
\end{equation}
for any $h\in\mathfrak{h}$ and $p\in\mathfrak{p}$. A decomposition of a real Lie algebra, such as $\mathfrak{su}(4)$ (the algebra of generators for $\mathrm{SU\left(4\right)}$), that satisfies equation \eqref{eq:Cartan_Decomp}, forms a Cartan decomposition. Furthermore, it involves an antiunitary operator $\Theta$ that divides $\mathfrak{g}$ as follows:
\begin{equation}
    \Theta h \Theta^{-1} = h, \quad \Theta p \Theta^{-1} =-p\;, \label{eq:antiunitary_cartan}
\end{equation}
such that \eqref{eq:Cartan_Decomp} and \eqref{eq:antiunitary_cartan} are equivalent.

Quantum Hamiltonians are closely related to Lie algebras and Lie groups. For instance, 
%the 2-qubit system is fully described by the real Lie algebra $\mathfrak{su}(4)$, which generates the group $\mathrm{SU\left(4\right)}$. In other words, 
the general two-qubit Hamiltonian \cite{Horodecki1995}:
\begin{equation}
    H_{2q} = t_0 \mathbbm{1}\otimes \mathbbm{1} + \sum_{i,j=1}^{3} a_i \sigma_{i}\otimes \mathbbm{1}+ b_j \mathbbm{1}\otimes\sigma_{j}+t_{ij}\sigma_{i}\otimes\sigma_{j} \label{eq:2qubitH}
\end{equation}
where $({\mathbbm{1}},\sigma_i )$ are $2\times 2$ Pauli matrices including the identity %operator, \bcomment{$\mathbbm{1}$ is the $2\times 2$ identity operator}
and $a_i,b_i,t_0,t_{ij}\in \mathbb{R}$, belongs to the algebra $\mathfrak{su}(4)$. The corresponding evolution operator $\exp(-iH_{2q}t)$  ($\hbar \equiv 1$) thus belongs to the Lie group $U\left(4\right)$ at any given time $t$. 

A known \cite{Khaneja2001,KHANEJA200111,Zhang2003} Cartan decomposition of $\mathfrak{su}(4)$ is, 
\begin{equation}
    \begin{gathered}
    \mathfrak{h}_2=  \operatorname{span}\left\{ i\sigma_{j}\otimes {\mathbbm{1}},\, i{\mathbbm{1}}\otimes \sigma_{k} \right\}\\
    \mathfrak{p}_2=  \operatorname{span}\left\{ i\sigma_{j}\otimes\sigma_{k}\right\}\\
    \mathfrak{g}_2 =   \mathfrak{h}_2\oplus \mathfrak{p}_2
    \label{eq:Cartan_2q}
\end{gathered}
\end{equation}
where $j,k\in\left\{x,y,z\right\} $ and $i=\sqrt{-1}$. The operator $\mathbbm{1}\otimes \mathbbm{1}$ is not in the basis, but it does belong to the algebra. The antiunitary operator $\Theta$ in \eqref{eq:antiunitary_cartan} is:
\begin{equation}
    \Theta=\left(\sigma_{y}\otimes\sigma_{y}\right)K \label{time reversal}
\end{equation}
where $K$ is the antiunitary complex conjugation operator on the computational basis $|jk\rangle$ with $j,k\in\left\{0,1\right\} $, namely
\begin{equation}
    K|jk\rangle=|jk\rangle,\quad KiK=-i.
\end{equation}
Physically, $\Theta$ is the time-reversal symmetry operator of a two-spin-1/2 system.

A Cartan decomposition splits 2-qubit Hamiltonians into entangling and non-entangling \cite{Khaneja2001,KHANEJA200111,Zhang2003}. This can be seen by analyzing the dynamics of entanglement using the Wootters' concurrence \eqref{eq:concurrence}. Any 2-qubit Hamiltonian $H$ that does not contain the identity, is an element of $\mathfrak{g}_2$. Consider a Hamiltonian $H_h\in i\mathfrak{h}_2$, as defined in equation \eqref{eq:Cartan_2q}. It has the general form:
\begin{equation}
    H_h=a_i \sigma_{i}\otimes {\mathbbm{1}} + b_j {\mathbbm{1}}\otimes \sigma_{j}\;,\label{H_h} 
\end{equation}
where $a_i,b_j\in \mathbb{R}$  and summation over repeating indices is implied. It can also be obtained by setting $t_{ij}\equiv 0$ in \eqref{eq:2qubitH}. For a state $\ket{\psi_0}$ with concurrence $C(0)$ that evolves in time with the Hamiltonian \eqref{H_h}: 
\begin{equation}
    C_\psi(t)=\left|\left\langle \psi_0\left|e^{iH_{h}t}\Theta e^{-iH_{h}t}\right|\psi_0\right\rangle \right|= C(0) \;. \label{eq:concurrence_in_time}
\end{equation}
where the last equality results from the Cartan identity \eqref{eq:antiunitary_cartan} that implies $e^{i H_h t}\Theta=\Theta e^{iH_h t}$. 

Let us denote by $H_p \in i\mathfrak{p}_2$:
\begin{equation}
    H_p=\sum_{i,j=1}^{3} t_{ij}\sigma_{i}\otimes\sigma_{j} \label{eq:hp}
\end{equation}
where $t_{ij} \in \mathbb{R}$. Substituting $H_p$ into \eqref{eq:concurrence_in_time} leads to $e^{iH_pt}\Theta=\Theta e^{-iH_pt}$, indicating, through a similar reasoning, that $C_\psi(t)$ may differ from $C(0)$. Note that \eqref{eq:antiunitary_cartan} suggests that \eqref{eq:concurrence_in_time} holds true regardless of the basis chosen. Consequently, we deduce that operators belonging to $\mathfrak{p}_2$ are able to entangle, whereas those in $\mathfrak{h}_2$ are not.

An alternative and potentially more intuitive explanation involves time-evolution operators related to $H_h$ and expressed as,
\begin{equation}
    U_h=e^{i(a_i \sigma_{i}\otimes {\mathbbm{1}}+ b_j {\mathbbm{1}}\otimes \sigma_{j})\,t}=e^{ia_i \sigma_{i}\,t}\otimes e^{itb_j \sigma_{j}\,t} \, \, .
\end{equation}
Thus, $U_h$ belongs to $\mathrm{SU\left(2\right)}\otimes\mathrm{SU\left(2\right)}$, indicating it acts independently on each qubit and does not alter their entanglement. It is evident that when $U_h$ operates on an initially separable state $\ket{A}\otimes\ket{B}$, it remains in a separable state. These insights illustrate that the entanglement characteristics of 2-qubit systems are dictated by the Cartan decomposition \eqref{eq:Cartan_Decomp}, where $\mathfrak{h}$ represents nonentangling operators, and $\mathfrak{p}$ represents entangling operators.

The 2-qubit Hamiltonian \eqref{eq:2qubitH} generates all operators in the group SU(4) using 15 real parameters. Comparison with Cartan decomposition \eqref{eq:Cartan_2q} suggests that 6 of them define the local (non-entangling) part of the evolution and the remaining 9 contribute to the non-local (entangling) part. The Cartan decomposition of the Lie group SU(4) \cite{Khaneja2001,KHANEJA200111,Zhang2003,Cirac2001} gives for any $U\in SU(4)$:
\begin{equation}
    U=K_2 \, e^{-i\left(\alpha\sigma_{x}\otimes\sigma_{x}+\beta\sigma_{y}\otimes\sigma_{y}+\gamma\sigma_{z}\otimes\sigma_{z}\right)} K_1 \label{eq:khaneja_evop}
\end{equation}
where $K_{1,2}\in SU(2)\otimes SU(2)$ (see Appendix \ref{app:khaneja}). Hence, any evolution operator can be decomposed into a single non-local operation preceded ($K_1$) and followed ($K_2$) by a local operation. This non-local part is characterized by 3 real parameters \footnote{Note that if $\alpha=\beta=\gamma=0$, then $U$ is local and does not entangle}, which generate a subspace of the whole space generated by $\mathfrak{p}_2$,
whose corresponding evolution operators belong to a subspace of the evolution operators allowed by  \eqref{eq:hp}. The additional information included in \eqref{eq:khaneja_evop} is that this subspace is equivalent, by local operations, to other subspaces of allowed evolution operators of \eqref{eq:hp}.

We have analyzed 2-qubit entanglement through the Wootters concurrence generated by the $\Theta$ operator. Uhlmann \cite{Uhlmann2000} has suggested a broader framework, indicating that various concurrence measures can be constructed using different antiunitary operators $\Theta$ as described in \eqref{eq:concurrence}, which also evaluate entanglement. Therefore, for 2-qubit systems, multiple formulations of concurrence can be derived. These formulations have associated Cartan decompositions \eqref{eq:Cartan_Decomp}, enabling us to define locality and ascertain the decomposition \eqref{eq:khaneja_evop} of evolution operators.

%Though we have focused on the 2-qubit concurrence in this section, Uhlmann \cite{Uhlmann2000} has shown that entanglement can be completely described by generalizations of the Wootters concurrence \eqref{eq:concurrence}. In general there could be more concurrence for a given system, each based on a different antiunitary operator, i.e. takes the place of $\Theta$ in equation \eqref{eq:concurrence}. This means for a general system one can use each definition of concurrence to create a Cartan decomposition \eqref{eq:Cartan_Decomp}, and thus define locality with respect to itself. In a similar manner, one can find the Cartan decomposition of the evolution operator \eqref{eq:khaneja_evop}.

%\comment{This should be more} Many of the results presented here can be generalized to other systems. This is discussed in section \ref{sec:generalizations}.

%%%%%%%%%%%%%%%%%%%%%%%%%%%%%%%%%%%%%%%%%%%%%%%%%%%%%%%%%%%%%%%%%%%%%%%%%%%%%
%%%%%%%%%%%%%%%%%%%%%%%%%%%%%%%%%%%%%%%%%%%%%%%%%%%%%%%%%%%%%%%%%%%
%%%%%%%%%%%%%%%%%%%%%%%%%%%%
%%%%%%%%%%%%%%%%%%%%%%%%%%%%%%%%%%%%%%%%%%%%%%%%%%%%%%%%%%%%%%%%%%%%%%%%%%%%%
%%%%%%%%%%%%%%%%%%%%%%%%%%%%%%%%%%%%%%%%%%%%%%%%%%%%%%%%%%%%%%%%%%%
%%%%%%%%%%%%%%%%%%%%%%%%%%%%%%%%%%%%%%%%%%%%%%%%%%%%%%%%%%%%%%%%%%%

\section{Antiunitary Symmetries and Tenfold Classification \label{sec:Topology}}
%%%%%%%%%%%%%%%%%%%%%%%%%%%%%%%%%%%%%%%%%%%%%%%%%%%%%%%%%%%%%%%%%%%%%%%%%%%%%
%%%%%%%%%%%%%%%%%%%%%%%%%%%%%%%%%%%%%%%%%%%%%%%%%%%%%%%%%%%%%%%%%%%
%%%%%%%%%%%%%%%%%%%%%%%%%%%%%%%%%%%%%%%%%%%%%%%%%%%%%%%%%%%%%%%%%%%

\subsection{Time-Reversal \texorpdfstring{$\Theta$}{Theta} and Particle-Hole \texorpdfstring{$\Pi$}{Pi} \label{sec:symmetries}}

Antiunitary symmetries play a central role in quantum entanglement, and as noted in Section \ref{sec:Cartan}, the Uhlmann concurrence \cite{Uhlmann2000} generalizes the Wootters concurrence using other antiunitary symmetries in addition to time-reversal. 
%to any system. This is done by allowing  more than a single concurrence and
For 2-qubit systems, a single antiunitary symmetry, time reversal as given in \eqref{time reversal}, defines concurrence. For two fermions, another antiunitary symmetry, particle-hole, has been proposed to account for a new kind of concurrence \cite{Schliemann2001,Eckert2002}. According to Wigner's theorem, any unitary quantum evolution can be given up to two nonequivalent antiunitary symmetries \cite{Moore}. 
 We can always choose them to be time-reversal $\Theta$ and particle-hole $\Pi$.
The exact forms of antiunitary symmetries are basis dependent, but square to $\pm \mathbbm{1}$ ($\mathbbm{1}$ is the identity operator) defining three basis-independent flavors, including non-existence. These features of quantum evolution result from the Lie algebra structure of the Hamiltonians of discrete systems. Hence, corresponding Lie algebras can be classified solely by means of these two antiunitaries and their squares. This is the essence of Cartan tenfold classification.

%%%%%%%%%%%%%%%%%%%%%%%%%%%%%%%%%%%%%%%%%%%%%%%%%%%%%%%%%%%%%%%%%%%%%%%%%%%%%
%%%%%%%%%%%%%%%%%%%%%%%%%%%%%%%%%%%%%%%%%%%%%%%%%%%%%%%%%%%%%%%%%%%
%%%%%%%%%%%%%%%%%%%%%%%%%%%%%%%%%%%%%%%%%%%%%%%%%%%%%%%%%%%%%%%%%%%
\subsection{Tenfold Classification \label{sec:tenfold}}

In condensed matter physics, Cartan classification is used to identify topological properties of quadratic Hamiltonians describing non (or weakly) interacting fermions. The identification of a Cartan symmetry class is formally equivalent to analyzing the topological properties of a Hamiltonian or its corresponding time evolution operator. If a Hamiltonian belongs to a subset of a Lie algebra generated by a Cartan decomposition \eqref{eq:Cartan_Decomp}, its evolution operator defines a symmetric space described by the tenfold classification \cite{Ludwig2016, Pires_Topology_book,Khaneja2001,Stone2010, Helgason1978, Caselle_Magnea_Review}. 

For a system lacking unitary symmetries, its symmetry class is determined by up to two possible antiunitary symmetries. By selecting time reversal $\Theta$ and particle-hole $\Pi$, as defined in section \ref{sec:symmetries} and characterized by their respective squares:
\begin{equation}
    \Theta^2=\pm \mathbbm{1},\quad \Pi^2=\pm \mathbbm{1},
\end{equation}
we obtain 3 "types" $(0,\pm 1)$ for each. Chirality, identified by the operator $S=\Theta\Pi$, is unitary but not a symmetry because it anticommutes with the Hamiltonian if it exists. The combination $(\Theta, \Pi, S)$ delineates 9 symmetry classes as indicated in the first four columns of table \ref{table:10fold_with_Ri}, along with the additional symmetry class AIII which is chiral despite its lack of antiunitary symmetries. Existing unitary symmetries allow block-diagonalizing a Hamiltonian. Each subsequent block can be assigned a symmetry class using the previous prescription. While these blocks may belong to the same symmetry class, it is not guaranteed. 
\begin{table}
    \centering
\begin{tabular}{|>{\centering}p{0.9cm}|>{\centering}p{0.5cm}|>{\centering}p{0.5cm}|>{\centering}p{0.5cm}|>{\centering}m{1.6cm}|>{\centering}m{1.6cm}|>{\centering}m{1.5cm}|}
\hline 
Class & $\Theta$ & $\Pi$ & $S$ & Stable $H$ Space & Evolution Operator Space & $\pi_{0}$ of stable $H$ Space\tabularnewline
\hline 
\hline 
A & 0 & 0 & 0 & $\mathcal{C}_{0}$ & $\mathcal{C}_{1}$ & $\mathbb{Z}$\tabularnewline
\hline 
AIII & 0 & 0 & $+$ & $\mathcal{C}_{1}$ & $\mathcal{C}_{0}$ & 0\tabularnewline
\hline 
AI & $+$ & 0 & 0 & $\mathcal{R}_{0}$ & $\mathcal{R}_{7}$ & $\mathbb{Z}$\tabularnewline
\hline 
BDI & $+$ & $+$ & $+$ & $\mathcal{R}_{1}$ & $\mathcal{R}_{0}$ & $\mathbb{Z}_{2}$\tabularnewline
\hline 
D & 0 & $+$ & 0 & $\mathcal{R}_{2}$ & $\mathcal{R}_{1}$ & $\mathbb{Z}_{2}$\tabularnewline
\hline 
DIII & $-$ & $+$ & $+$ & $\mathcal{R}_{3}$ & $\mathcal{R}_{2}$ & 0\tabularnewline
\hline 
AII & $-$ & 0 & 0 & $\mathcal{R}_{4}$ & $\mathcal{R}_{3}$ & 2$\mathbb{Z}$\tabularnewline
\hline 
CII & $-$ & $-$ & $+$ & $\mathcal{R}_{5}$ & $\mathcal{R}_{4}$ & 0\tabularnewline
\hline 
C & 0 & $-$ & 0 & $\mathcal{R}_{6}$ & $\mathcal{R}_{5}$ & 0\tabularnewline
\hline 
CI & $+$ & $-$ & $+$ & $\mathcal{R}_{7}$ & $\mathcal{R}_{6}$ & 0\tabularnewline
\hline 
\end{tabular}
	\caption{The tenfold classification of quadratic Hamiltonians. The symmetry class is determined by the 2 antiunitary symmetries, time reversal $\Theta$ and particle-hole $\Pi$ and chirality $S$ as displayed in corresponding columns where $0$ means no symmetry of that kind, "+" ("-") means the symmetry operator squares to $+\mathbbm{1}$ ($-\mathbbm{1}$). The fourth and fifth columns indicate the label of the corresponding symmetric spaces. The last column displays the $\pi_0$ homotopy group of the Hamiltonian space which provides possible  topological numbers of a $0$-dimensional system.
 }
	\label{table:10fold_with_Ri}
\end{table}

%%%%%%%%%%%%%%%%%%%%%%%%%%%%%%%%%%%%%%%%%%%%%%%%%%%%%%%%%%%%%%%%%%%%%%%%%%%%%
%%%%%%%%%%%%%%%%%%%%%%%%%%%%%%%%%%%%%%%%%%%%%%%%%%%%%%%%%%%%%%%%%%

We identify two possible symmetric spaces. One is associated to Hamiltonians (Lie algebras) and the other is associated with unitary evolution operators (Lie groups). Quadratic Hamiltonians in condensed matter are often described by large $n\times n$ Hamiltonian matrices (e.g. tight-binding models). In the limit $n\rightarrow \infty$,  Hamiltonian symmetric spaces, also called stable, are denoted $\mathcal{C}_{0,1}$ for classes A and AIII, and $\mathcal{R}_i$ for the remaining classes, with $i=\{0,1,2,3,4,5,6,7\}$. 

%A given system has two associated symmetric spaces, space of Hamiltonians and of evolution operators. For a very large system (also called "stable", see next section for more detail) there is a naming scheme for the symmetric space of Hamiltonians: $\mathcal{C}_{0,1}$ for classes A and AIII, and $\mathcal{R}_i$ for the rest, with $i=\{0,1,2,3,4,5,6,7\}$. This is presented in column 5 in table \ref{table:10fold_with_Ri}. If a Hamiltonian possesses unitary symmetries, then after it is block-diagonalized using them, each block defines a symmetric space by itself using its antiunitary symmetries. 

The expression $U(t)=\exp{(-iH t)}$ for the connection between Hamiltonians $H$ and their corresponding time evolution operators illustrates the analogous relationship between Lie groups and Lie algebras, which is also observed in symmetric spaces. A key aspect in the Cartan classification of these symmetric spaces is their topological characteristics, which arise due to Bott periodicity \cite{Milnor}. This periodicity dictates that Hamiltonians associated with a symmetric space $\mathcal{R}_i$ correspond to time evolution operators linked to $\mathcal{R}_{i-1}$, following a periodicity of modulo 8, such that $\mathcal{R}_8 \equiv \mathcal{R}_0$. This relationship can be schematically depicted by:
\begin{equation}
    H \in \mathcal{R}_j \Leftrightarrow U(t) = e^{-iHt} \in \mathcal{R}_{j+1},
\end{equation}
and a similar pattern applies to $\mathcal{C}_{0,1}$ with modulo 2 periodicity. This relation between symmetric spaces is a consequence of the Cartan decomposition. If $\mathcal{R}_j$ is the space generated for $\mathfrak{g}$ by \eqref{eq:Cartan_Decomp}, then $\mathcal{R}_{j+1}$ is obtained using the decomposition of $\mathfrak{h}$ \cite{Stone2010}. Another consequence of Bott periodicity is the topological transcription: 
\begin{equation}
    \pi_i(\mathcal{R}_j)=\pi_{i-1}(\mathcal{R}_{j+1}) \label{eq:pii_Ri}
\end{equation}
of the previous relation, where $\pi_i({R}_j)$ is the $i$-th homotopy group of the symmetric space ${R}_j$ \cite{Nakahara1990,Milnor,Pires_Topology_book}, where indices $i,j$ are taken modulo 8. Thus, to a lower-order topology for $\mathcal{R}_{j+1}$ corresponds a higher-order one for $\mathcal{R}_{j}$, in the ladder-like structure \eqref{eq:pii_Ri}. This structure hints at the particular role played by the homotopy $\pi_0$, which accounts for the connectedness of symmetric spaces, as is known for $O(3)$ and $SO(3)$ groups. 

These results have also been considered in the contexts of scattering theory \cite{Caselle_Magnea_Review} and transfer matrices \cite{Mudry2001}. In Section \ref{sec:Hamiltonian_symmetric_space} we discuss how to use \eqref{eq:pii_Ri} to calculate topological invariants, and physical implications are discussed in Section \ref{sec:2qubits}. 

Recognizing symmetric spaces of Hamiltonians and time evolution operators can be challenging. We will now elaborate on the common approaches to this task, aiming to shed light on the physical significance of these occasionally ambiguous concepts. Section \ref{sec:2qubits} will apply this discussion to pinpoint the topological characteristics of 2-qubit systems and their connection to entanglement.

%%%%%%%%%%%%%%%%%%%%%%%%%%%%%%%%%%%%%%%%%%%%%%
%%%%%%%%%%%%%%%%%%%%%%%%%%%%%%%%%%%%%%%%%%%%%%
%%%%%%%%%%%%%%%%%%%%%%%%%%%%%%%%%%%%%%%%%%%%%%

\subsection{Symmetric Spaces for Evolution Operators \label{sec:Evolution_symmetric_space}}
A symmetry class is identified by finding the symmetric space corresponding to its evolution operators. This consists of describing the evolution operator as a Lie group or one of its subsets. 

As a starting example, consider a $n$-level system without symmetries (either unitary or antiunitary) and denote by $H_\mathrm{ns}$ this "no symmetry" Hamiltonian. %, where "ns" stands for "no symmetry". 
The time evolution operators $\exp{(-iH_\mathrm{ns} t)}$ are $n\times n$ unitary matrices. Since there is no symmetry restriction, these operators span the entire group $U(n)$, known as the symmetric space $C_1(n)$ assigned to symmetry class A \cite{Schnyder2009,Caselle_Magnea_Review,Ludwig2016} (see Table \ref{table:10fold_with_Ri}). This is a fundamental example in quantum mechanics where symmetric spaces naturally appear.

%\comment{We will not use notations prevalent in other references e.g. \cite{Schnyder2008, Kitaev2009, Stone2010} where these are called the "Space of Hamiltonians" or "Classifying space" of class A. Should I write this disclaimer? \cite{Schnyder2008,Schnyder2009} specifically is quite misleading}

Consider now Hamiltonians that possess antiunitary symmetries, hence the space of time evolution operators becomes a subspace of $U(n)$. For example, consider time-reversal $\Theta$-symmetric Hamiltonians with $\Theta^2=+\mathbbm{1}$, i.e. class AI. In this case, time evolution operators are restricted by symmetry and are described by $U(n)/O(n)\equiv R_7(n)$ \cite{Stone2010,Ludwig2016}, where $O(n)$ is the group of orthogonal $n\times n$ matrices. In other words, it is the space of unitary matrices, where we identify any two matrices that differ by an orthogonal transformation. This  important result will be applied to $n=4$ in Section \ref{sec:2qubits}. 

To illustrate the difference between Lie algebras, symmetric spaces and their topology, it is of interest to consider $SU(n)/SO(n)$ sometimes used to represent $R_7(n)$ instead of $U(n)/O(n)$ \cite{Caselle_Magnea_Review}. This is because the algebras corresponding to those Lie groups are identical apart from 
the $n\times n$ unit matrix which is included as a generator or not. This addition or omission does not change the physics as it only adds a (time-dependent) total phase to all states. However, it bears notable consequences concerning topological properties because for any $n \geq 3$:
\begin{equation}
    \pi_1 (U(n)/O(n))=\mathbb{Z} \neq 0= \pi_1 (SU(n)/SO(n)). \label{eq:u/oneqsu/so}
\end{equation}
Given that these are mathematically legitimate symmetric spaces, it is essential to consider their distinctions within the physical context \footnote{ Another version of $R_7(4)$ gives $\pi_1 (R_7 (n))=\mathbb{Z}_n$.}. The importance of a careful construction of symmetric spaces is further discussed in Section \ref{sec:finite}.

%%%%%%%%%%%%%%%%%%%%%%%%%%%%%%%%%%%%
%%%%%%%%%%%%%%%%%%%%%%%%%%%%%%%%%%%%
%%%%%%%%%%%%%%%%%%%%%%%%%%%%%%%%%%%%

\subsection{Hamiltonian Symmetric Space \label{sec:Hamiltonian_symmetric_space}}
Symmetry classes are determined from the symmetric space linked to a set of Hamiltonians. In Section \ref{sec:tenfold}, we explored how symmetric spaces related to both an evolution operator and its Hamiltonian \cite{Stone2010,Ludwig2016} can be connected. In this section, we apply this connection to systems lacking unitary symmetries.

Two primary aspects in identifying the symmetric space of a set of Hamiltonians include their antiunitary symmetries and the classification of eigenenergies based on their signs. The classification is derived from flattened Hamiltonians, where energies are designated as $\pm 1$ depending on whether they are positive or negative. Then, antiunitary symmetries are identified and the flattened Hamiltonian is correspondingly modified so as to contain any operator abiding by them. Finally, resulting flattened Hamiltonians are assigned a Lie algebra. The coefficients given to each basis element of that algebra form a vector - the Hamiltonian is represented by a vector in an operator space. The resulting symmetric space is obtained by identifying Hamiltonians that can be continuously deformed into each other without breaking a symmetry or changing the sign of eigenvalues \cite{Kitaev2009,Pires_Topology_book}. The latter condition 
%on the sign of the eigenvalues is analogous to the necessity of 
amounts to requesting the existence of an energy gap in the spectrum (also known as the "no band crossing" rule in condensed matter) \cite{Schnyder2008,Schnyder2009,BernevigBook2013,Ludwig2016,Chiu2016}. 

Back to our previous example, the symmetric space of Hamiltonians $H_{ns}$ for a $n$-level quantum system with no symmetries is the set of Grassmannians over $\mathbb{C}^n$ \cite{Kitaev2009,Morimoto2015,Ludwig2016,Pires_Topology_book}:
\begin{equation}
    C_0 (n) = \bigsqcup_{0\leq k\leq n} U\left(n\right)/\left(U\left(n-k\right)\times U\left(k\right)\right) \, . \label{complex_grassman}
\end{equation}
 $C_0 (n)$ is a disjoint union of $n +1$ quotient-spaces of the unitary group. The index $k$ counts  negative eigenvalues of $H_{ns}$, such that Hamiltonians having distinct $k$ cannot be continuously deformed into one another. This property provides a natural framework for the understanding of the  topology associated with symmetric spaces in zero-dimensional systems. Topologically distinct Hamiltonians belong to different path components, i.e. disconnected subspaces of their respective symmetric space, either $ C_i (n)$ or $R_i (n)$ \cite{Mudry2001}. The set of different path components is denoted $\pi_0(C_i (n))$ ($\pi_0(R_i (n))$ respectively). For a $n$-level Hamiltonian $H_{ns}$ without symmetry, the set $\pi_0(C_0(n))$ assigns an integer, a "topological invariant" of the system to each topologically distinct component of $ C_0 (n)$. In the limit $n\rightarrow\infty$, $C_0(n)$ tends to the stable symmetric space $\mathcal{C}_0$, and $\pi_0 (\mathcal{C}_0)$ is given by the integers $\mathbb{Z}$, where each integer corresponds to a numerical imbalance between positive and negative eigenvalues. 
A similar analysis holds for all other symmetric spaces and is summarized in the last column of table \ref{table:10fold_with_Ri}. We discuss the case of finite matrices of size $n$ in section \ref{sec:finite}.

%For a $d$-dimensional system, the Hamiltonian is no longer a matrix, but a collection of matrices which depend on $d$ parameters. 
For a spatially $d$-dimensional physical system, the Hamiltonian matrix depends on $d$  parameters so as to define a $d$-dimensional set of matrices, namely a set of points in the symmetric space instead of a single point. Topologically distinct Hamiltonians still cannot be continuously deformed into one another, e.g %, i.e. deformations of the aforementioned set.
%As before, topologically distinct Hamiltonians cannot be deformed continuously into one another, but instead of a point in the symmetric space the Hamiltonian is now a set of points. 
  class AI infinitely large Hamiltonians defined on the $d$-sphere are topologically described by a map from the $d$-sphere to $\mathcal{R}_0$. The number of such distinct maps is the $d$-homotopy group of $\mathcal{R}_0$, denoted $\pi_d(\mathcal{R}_0)$, which sets  the corresponding topological number \cite{Stone2010}.

For the noteworthy instance of $d$-dimensional crystals with translational symmetry, the Bloch Hamiltonian forms a relation between the Brillouin zone, defined by the wavevector $\vec k$, and the symmetric space.
As a first approximation, one can consider the Brillouin zone as the sphere $S^d$. Subsequent refinements involve considering "weak topology," which is beyond the reach of the tenfold classification \cite{Avron1983,Kitaev2009,Ertem2017,Schnyder2009}.
As anticipated, topological invariants are represented by $\pi_d(\mathcal{R}_i)$ (or $\pi_d(\mathcal{C}_i)$). It is essential to note, however, that the Brillouin zone, serving as the base space, is inverted by antiunitary symmetries, since it is a Fourier transform of real space. This property changes the homotopy groups and the correct topological numbers are provided by $\pi_{-d}(\mathcal{R}_i)$ (or $\pi_{-d}(\mathcal{C}_i)$) with "$-d$" calculated modulo 8 \cite{Stone2010}. To illustrate this point, consider class AI stable ($n\rightarrow\infty$), 0-dimensional Hamiltonians, characterized by time reversal $\Theta$ antiunitary symmetry with $\Theta^2=+\mathbbm{1}$. These Hamiltonians and their evolution operators have symmetric spaces $\mathcal{R}_0$ and $\mathcal{R}_7$ respectively, with corresponding (Chern) integers given by $\pi_0(\mathcal{R}_0)=\mathbb{Z}$.  According to \eqref{eq:pii_Ri},
\begin{equation}
    \pi_1(\mathcal{R}_7)=\pi_0(\mathcal{R}_0)=\mathbb{Z}. \label{eq:pi1R7=pi0R0}
\end{equation}
The first homotopy $\pi_1$ of a symmetric space quantifies closed paths (loops) in that space which cannot be continuously deformed into one another. Here, the time $t$ parameterizes loops of the time evolution operators, that is, the base space includes only periodic time evolution. 

Differences between closed loops in the Hilbert space result in geometrical (Berry) phases \cite{Berry1984, Simon1983, Aharonov1987}. The notion of a geometrical phase has also been extended to closed loops in time for evolution operators \cite{Uhlmann1986}. Our findings imply that Berry phases capture the topological characteristics of the space of the time-evolution operator via closed loops, particularly $\pi_1(\mathcal{R}_7)$. As indicated in \eqref{eq:pi1R7=pi0R0}, this means the Berry phase serves as a measure for topological invariants of 0-dimensional Hamiltonians in class AI. Given that \eqref{eq:pii_Ri} is not limited to any specific symmetry class, it suggests that the Berry phase evaluates topological invariants for 0-dimensional Hamiltonians across all classes.

%%%%%%%%%%%%%%%%%%%%%%%%%%%%%%%%%%%%%%%%%%%%%%%%%
%%%%%%%%%%%%%%%%%%%%%%%%%%%%%%%%%%%%%%%%%%%%%%%%%

\subsection{Stable Topological Features \label{sec:finite}}
In condensed matter systems, the limit $n\rightarrow\infty$ for the number of energy levels of the associated Hamiltonians is often cited. \cite{Kitaev2009,Ludwig2016,Chiu2016,Caselle_Magnea_Review,Pires_Topology_book,Stone2010}, occasionally in extreme situations where $n$ is merely 2. Frequently, few-level systems serve as an approximation for multi-level systems by neglecting slowly changing or constant degrees of freedom. For example, when constructing a tight-binding Hamiltonian, the majority of atomic orbitals are usually excluded. The assumption is that introducing more levels does not change the symmetry class, thus supporting the limit $n\rightarrow\infty$. Mathematically, it is known that for significantly large $n$, the lower homotopy groups, indicated by $\pi_i$, reach stability and cease to vary with $n$; this is known as homotopy stabilization. Additionally, the computation of homotopy groups is simplified as $n$ tends to infinity. \cite{Pires_Topology_book,Stone2010}.

%In our case this does not hold. The case of two distinguishable qubits is $n=4$, thus effects of finite $n$ may be observed. 

For finite values of \( n \), homotopy groups display increased complexity compared to the scenario where \( n \) tends toward infinity. This complexity becomes evident through the analysis discussed in the preceding section \ref{sec:Hamiltonian_symmetric_space}, which considers a straightforward \( n \)-level system without symmetries. As described in \cite{Morimoto2015}, under these conditions, the space \eqref{complex_grassman} is comprised of \( (n+1) \) distinct disconnected components. From a physical standpoint, this indicates the existence of \( (n+1) \) unique topological phases, each characterized by an integer, resulting in a \(\mathbb{Z}_{n+1}\) topology. As $n$ approaches infinity, the space $C_0(n)$ evolves into $\mathcal{C}_0$, which is the stable space associated with the same symmetry class, incorporating a $\mathbb{Z}$ topology. According to the stabilization hypothesis, when $n$ is sufficiently large, a finite space turns topologically non-trivial if the corresponding stable space is also non-trivial. It has been suggested that this value of $n$ is generally quite low (approximately 4) concerning the low homotopy groups $\pi_0$ and $\pi_1$ \cite{Milnor}.

In other words, the space $C_0(n)$ as described in \eqref{complex_grassman} represents an "unstable" form of $\mathcal{C}_0$. It is not unique, meaning there are alternative spaces that converge to $\mathcal{C}_0$. One illustrative example is the space $\tilde{C}_0(n) = U\left(2n\right)/\left(U\left(n\right)\times U\left(n\right)\right) \times \mathbb{Z}$, which approaches $\mathcal{C}_0$ as $n$ approaches infinity. It should be highlighted that these spaces may have distinct homotopy groups. Consequently, constructing such finite spaces demands a thorough consideration of the physical requirements, as explained, for instance, in section \ref{sec:2qubits}. Similar arguments for other stable spaces are shown in  Table \ref{tab:sym_space_finite_approx}.

Another aspect to consider for finite systems is the applicability of Bott periodicity \eqref{eq:pii_Ri}. For instance, in section \ref{sec:2qubits}, we evaluate $\pi_0 (R_0 (4))=\mathbb{Z}_5$. By comparing this result with the three versions of $R_7(4)$ provided in Table \ref{tab:sym_space_finite_approx}, it appears that Bott periodicity \cite{Milnor} fails for $n=4$. Nevertheless, with one exception, the majority of the homotopy groups shown converge to $\mathbb{Z}$ as $n$ tends to infinity, corresponding to the homotopy group of the stable space $\mathcal{R}_7$. Therefore, it can be asserted that the equation \eqref{eq:pi1R7=pi0R0} holds true more generally, implying that Bott periodicity holds even for small $n$. In terms of physics, Bott periodicity \eqref{eq:pii_Ri} suggests that the Hamiltonian and the evolution operator describe the same physical phenomena, applicable for finite $n$. Hence, we suggest that if a Hamiltonian demonstrates a non-trivial topological property, the corresponding evolution operator must also exhibit this non-trivial topological characteristic, and vice versa.
 
\renewcommand{\arraystretch}{1.2} %modify height of rows
\begin{table}
\centering
\resizebox{\linewidth}{!}{%
\begin{tabular}{|c|c|c|c|c|}
\hline 
Class & Name & Space & $\pi_{0}$ & $\pi_{1}$\tabularnewline
\hline 
\hline 
\multirow{3}{50pt}{
\centering
$H:A$  
$U\left(t\right):AIII$
} & \multirow{3}{*}{$C_{0}\left(n\right)$} & $\bigsqcup_{k=0}^{n}U\left(n\right)/\left(U\left(n-k\right)\times U\left(k\right)\right)$ & $\mathbb{Z}_{n+1}$ & $0$\tabularnewline
\cline{3-5}
 &  & $\mathbb{Z}\times U\left(n\right)/\left(U\left(n/2\right)\times U\left(n/2\right)\right)$ & $\mathbb{Z}$ & $0$\tabularnewline
\cline{3-5}
 &  & $\mathbb{Z}\times SU\left(n\right)/\left(SU\left(n/2\right)\times SU\left(n/2\right)\right)$ & $\mathbb{Z}$ & 0\tabularnewline
\hline 
\multirow{3}{50pt}{
\centering
$H:AI$  $U\left(t\right):BDI$
} & \multirow{3}{*}{$R_{0}\left(n\right)$} & $\bigsqcup_{k=0}^{n}O\left(n\right)/\left(O\left(n-k\right)\times O\left(k\right)\right)$ & $\mathbb{Z}_{n+1}$ & $\mathbb{Z}_{2}$\tabularnewline
\cline{3-5}
 &  & $\mathbb{Z}\times O\left(n\right)/\left(O\left(n/2\right)\times O\left(n/2\right)\right)$ & $\mathbb{Z}$ & $\mathbb{Z}_{2}$\tabularnewline
\cline{3-5}
  &  & $\mathbb{Z}\times SO\left(n\right)/\left(SO\left(n/2\right)\times SO\left(n/2\right)\right)$ & $\mathbb{Z}$ & $0$\tabularnewline
\hline 
\multirow{3}{45pt}{
\centering
$H:CI$
$U\left(t\right):AI$
} & \multirow{3}{*}{$R_{7}\left(n\right)$} & $U\left(n\right)/O\left(n\right)$ & $0$ & $\mathbb{Z}$\tabularnewline
\cline{3-5}
 &  & $SU\left(n\right)/SO\left(n\right)$ & $0$ & $0$\tabularnewline
\cline{3-5}
 &  & $U\left(n\right)/SO\left(n\right)$ & $0$ & $\mathbb{Z}$\tabularnewline
\hline 
\end{tabular}}

\caption{Low homotopy groups for some finite symmetric spaces. The $\pi_1$ column applies for $n\geq 3$ in the case of $R_7(n)$, and for $n\geq4$, with $3\leq k\leq n-k$, for $R_0(n)$.}
\label{tab:sym_space_finite_approx}
\end{table}
\renewcommand{\arraystretch}{1} %make the height regualr again for later tables

%\comment{There is already a lot of wordy text. I suggest to remove the next paragraph}

%\bcomment{Physically, it is possible that a system has a finite number of different phases but the Berry phase is unlimited. For example, a single Aharonov-Bohm flux is a single point defect on a 2-d plane, which is geometrically independent of the intensity. From the electronic point of view there are many phases depending on the intensity of the flux. In the same manner, the Berry phase can wind many times depending on the absolute physical energy scale where the Hamiltonian phase topological transition depends only on relative energy scales.}

%%%%%%%%%%%%%%%%%%%%%%%%%%%%%%%%%%%%%%%%%%%%%%%%%%%%%%%%%%%%%%%%%%%%%%%%%%%%%
%%%%%%%%%%%%%%%%%%%%%%%%%%%%%%%%%%%%%%%%%%%%%%%%%%%%%%%%%%%%%%%%%%%
%%%%%%%%%%%%%%%%%%%%%%%%%%%%%%%%%%%%%%%%%%%%%%%%%%%%%%%%%%%%%%%%%%%
%%%%%%%%%%%%%%%%%%%%%%%%%%%%%%%%%%%%%%%%%%%%%%%%%%%%%%%%%%%%%%%%%%%%%%%%%%%%%
%%%%%%%%%%%%%%%%%%%%%%%%%%%%%%%%%%%%%%%%%%%%%%%%%%%%%%%%%%%%%%%%%%%
%%%%%%%%%%%%%%%%%%%%%%%%%%%%%%%%%%%%%%%%%%%%%%%%%%%%%%%%%%%%%%%%%%%
\section{Two qubits and the Tenfold Classification \label{sec:2qubits}}
In this part, we suggest that 2-qubit systems can be topologically characterized by the Cartan tenfold classification scheme. Next, we precisely define the symmetric spaces associated with both the evolution and Hamiltonian operators on their own. This method will uncover different aspects of the relationship between entanglement and topological properties. From the beginning, it is important to note that the Hamiltonian \eqref{eq:2qubitH} for a 2-qubit system does not possess symmetries. Consequently, it corresponds to our previous example of an $n$-level system devoid of symmetries, classified as class A. Since the system lacks any spatial or continuous components, the spatial dimension $d$ for this model is $0$.

Abiding time reversal symmetry \eqref{time reversal}, with $\Theta^2 =+ \mathbbm{1}$, leads to  \eqref{eq:hp} for the Hamiltonian $H_p$. The absence of particle-hole symmetry and other unitary symmetries implies that $H_p$ is in the AI class. The Hamiltonian $H_p$ describes two interacting spin-$1/2$ particles without an applied external magnetic field. In cases where magnetic fields are present, they are described by the operators $i \, \mathfrak{h}_2$ \eqref{eq:Cartan_2q}. Incorporating them into the Hamiltonian breaks time reversal symmetry and shifts the system to class A.

Using the Lie group structure of the evolution operator allows finding the corresponding symmetric space. This method illuminates the direct relationship between operator locality, as discussed in Section \ref{sec:Cartan}, and the symmetry class. The time evolution operator $U_g$ for a 2-qubit Hamiltonian \eqref{eq:2qubitH} is a 4$\times$4 unitary matrix, i.e. $U_g \in U(4)$. % included identity 
For time-reversal invariant systems, the Hamiltonian $H_p$ in \eqref{eq:hp} does not include  $\mathfrak{h}$ terms as defined in \eqref{eq:Cartan_2q}, which generate a single qubit, representing local operations characterized by $SU(2)\otimes SU(2)$, which is locally isomorphic to SO(4)  \cite{zee2016group}. The observation that local operations form the orthogonal group is discussed in \cite{Makhlin2000,Eckert2002}.

To obtain the symmetric space for the evolution operators $U_p$, we regard operators that differ by local transformations only as equivalent. As a result, the symmetric space is $U(4)$ (governed by $\mathfrak{g}$) modulo single-qubit operations (governed by $\mathfrak{h}$): 
\begin{equation}
    U_p (t)\equiv \exp{(iH_p t)} \in U(4)/SO(4) \equiv R_7 (4), \label{eq:evolution_symmetric_space}
\end{equation} 
associated with class AI. Relation \eqref{eq:evolution_symmetric_space} has thus been given a physical meaning as defining the space of entangling operators \cite{Khaneja2001,KHANEJA200111,Zhang2003}. As detailed in section \ref{sec:Evolution_symmetric_space} and table \ref{tab:sym_space_finite_approx}, this space is topologically non-trivial:
\begin{equation}
    \pi_1(U(4)/SO(4))=\mathbb{Z} \, .
\end{equation}
Revisiting the earlier discussed distinctions between different definitions of $R_7(4)$ in Sections \ref{sec:Evolution_symmetric_space} and \ref{sec:finite}, it was noted that local operations were represented by $SU(2)\otimes SU(2) \cong SO(4)$, which does not include multiplying a state by a global overall phase. As a total and global phase does not induce entanglement, we have the possibility to add the identity operator to the generators of $U_h$, resulting in $U_h\in SO(4)\times U(1)$. In this framework, $U_p (t) \in SU(4)/SO(4)$, so that it is associated to class AI \cite{Ludwig2016,Stone2010}. Note that this space is topologically trivial, as $\pi_1(SU(4)/SO(4))=0$, as shown in Table \ref{tab:sym_space_finite_approx}.
The difference between the two outcomes becomes clear when examining the physical interpretation of $\pi_1(U(t))$ as a Berry phase, as elaborated in section \ref{sec:Hamiltonian_symmetric_space}. As a result, when the overall phase is removed from the symmetric space, the topological numbers disappear. Nonetheless, as will be demonstrated, this disappearance does not occur in the Hamiltonian, implying that the discrepancy stems from varied measurement techniques instead of an inherent change in topology.

The symmetric space for two-qubit Hamiltonians is characterized by a 4-level system without  symmetries, as detailed in \eqref{eq:2qubitH} and further elaborated in Section \ref{sec:Hamiltonian_symmetric_space}. Consequently, it corresponds to \eqref{complex_grassman} for $n=4$, classifying the two-qubit system within class A. As detailed in Section \ref{sec:finite}, this system presents a nontrivial topology, and topological invariants are expressed as:
\begin{equation}
    \pi_0(C_0 (4))=\mathbb{Z}_5. \label{eq:homotopy_grassman}
\end{equation}
Consider $H_p$, as delineated in \eqref{eq:hp}. It is a 2-qubit Hamiltonian that maintains time-reversal symmetry but does not exhibit other symmetries. The Hamiltonian $H_p$ incorporates all symmetry-allowed operators and lacks any unitary symmetry. To determine the corresponding symmetric space \cite{Pires_Topology_book}, consider the operation of time-reversal symmetry within the Bell basis, denoted by $\Theta_B = K$ \cite{Wootters1998}. Hence, as shown in \eqref{eq:antiunitary_cartan}, the Hamiltonian $H_p$ commutes with the complex conjugation operator $K$, implying that $H_p ^*=H_p $ is a real matrix when expressed in the Bell basis. Additionally, being Hermitian, $H_p$ is both real and symmetric, allowing it to be orthogonally diagonalized.The flattened Hamiltonian is
\begin{equation}
    H_{p,\,\mathrm{flat}} = O\left(\begin{array}{cc}
\mathbbm{1}_{4-k} & 0\\
0 & -\mathbbm{1}_{k}
\end{array}\right)O^{T} \label{eq:Hflat}
\end{equation}
where $O\in O(4)$, the group of $4\times 4$ orthogonal matrices, $\mathbbm{1}_N$ is the $N\times N$ identity matrix and $k\in \left\{0,1,2,3,4\right\} $ is the number of negative eigenvalues of $H_p$. Any two flattened Hamiltonians with the same diagonal form are topologically equivalent, as argued in section \ref{sec:Hamiltonian_symmetric_space}. We thus identify any two operators related by the transformation:
\begin{equation}
    O\rightarrow O\left(\begin{array}{cc}
V_{4-k} & 0\\
0 & W_{k}
\end{array}\right)
\end{equation}
where $V_{4-k}\in O(4-k)$ and $W_{k}\in O(k)$. Since $H_p$ does not admit other symmetries, the space of operators $O$ fully describes $H_{p,\,\mathrm{flat}}$. Therefore, the space of topologically distinct Hamiltonians with $k$ negative eigenvalues is the real Grassmannian manifold,
\begin{equation}
   G\left(k,\mathbb{R}^4 \right) = O\left(4\right)/\left(O\left(4-k\right)\times O\left(k\right)\right) \, .
\end{equation}
In a system characterized by the Hamiltonian $H_p$, the spectrum of potential negative eigenvalues varies between 0 and 4, thereby defining the space of inequivalent Hamiltonians as
\begin{equation}
    R_0 (4) =\bigsqcup_{0\leq k\leq4}O\left(4\right)/\left(O\left(4-k\right)\times O\left(k\right)\right) \label{eq:R_0 finite}
\end{equation}
which implies that physical systems displaying entanglement belong to class AI.

While this derivation of symmetric spaces is based on particular characteristics of two-qubit systems, such as demonstrating local operations and flattened Hamiltonians with orthogonal matrices, these considerations are broadly applicable and can be expanded to accommodate any $n$-level system \cite{Eckert2002}.
Topological considerations  of \eqref{eq:R_0 finite} similar to those of complex Grassmannians \eqref{complex_grassman} lead to
\begin{equation}
    \pi_0(R_0 (4))=\mathbb{Z}_5. \label{eq:pi0R04}
\end{equation}
%This is again a finite approximation of $\pi_0(\mathcal{R}_0)=\mathbb{Z}$ for $n\rightarrow\infty$. 
Adding to $H_p$ terms in $i \, \mathfrak{h}$ breaks time reversal symmetry and removes the reality condition that leads to \eqref{eq:Hflat}. The diagonalizing matrices are now unitary, and the symmetric space is a complex Grassmannian  \eqref{complex_grassman} whose  topological numbers are given by \eqref{eq:homotopy_grassman}. 
%(the proof is exactly as in \cite{Kitaev2009,Pires_Topology_book}). As explained before, this space has a $\mathbb{Z}_5$ topological invariant. 

In conclusion, 2-qubit systems inherently exhibit non-trivial topological properties both when time-reversal symmetry is present and when it is absent. They encompass the set $\mathbb{Z}_5$ of topological invariants, indicating the existence of 5 distinct topological phases. The phases for systems with time-reversal symmetry are computed and interpreted in Section \ref{sec:topo_phases}.
The findings for the 2-qubit system provide a physical insight into the issue described in \eqref{eq:pii_Ri} within section \ref{sec:finite}. For a 2-qubit system with time-reversal symmetry, \eqref{eq:pi1R7=pi0R0} is not directly applicable due to the finite size of the Hamiltonian. Specifically, $R_7(4)$ in \eqref{eq:evolution_symmetric_space} represents a topologically non-trivial space with $\pi_1(R_7(4))=\mathbb{Z}$, which is distinct from $\pi_0(R_0 (4))=\mathbb{Z}_5$. Analogous results are relevant for 2-qubit systems lacking time-reversal symmetry by substituting $R_0(4)$ and $R_7(4)$ with $C_0(4)$ and $C_1(4)$, respectively. 

The presence of nontrivial $\pi_1$ homotopies suggests the occurrence of topological phase transitions that can be explored through a Berry phase. However, these transitions do not always align directly with the ones presented in Section \ref{sec:Topological_Invariants}. Importantly, since nontrivial homotopies are also present in time-reversal symmetric Hamiltonians and evolution operators, the non-zero Berry phase is produced not only by magnetic fields but also by entangling operators. Indeed, it can be shown that the resulting Berry phase is a function of the concurrence \eqref{eq:concurrence}, implying that it can be measured through an interference experiment.

%%%%%%%%%%%%%%%%%%%%%%%%%%%%%%%%%%%%%%%%%%%%%%%%%%%%%%%%%%%%%%%%%%%%%%%%%%%%%
%%%%%%%%%%%%%%%%%%%%%%%%%%%%%%%%%%%%%%%%%%%%%%%%%%%%%%%%%%%%%%%%%%%
%%%%%%%%%%%%%%%%%%%%%%%%%%%%%%%%%%%%%%%%%%%%%%%%%%%%%%%%%%%%%%%%%%%
%%%%%%%%%%%%%%%%%%%%%%%%%%%%%%%%%%%%%%%%%%%%%%%%%%%%%%%%%%%%%%%%%%%%%%%%%%%%%
%%%%%%%%%%%%%%%%%%%%%%%%%%%%%%%%%%%%%%%%%%%%%%%%%%%%%%%%%%%%%%%%%%%
%%%%%%%%%%%%%%%%%%%%%%%%%%%%%%%%%%%%%%%%%%%%%%%%%%%%%%%%%%%%%%%%%%%

\section{Topological phases of 2-qubits with time reversal symmetry \label{sec:topo_phases}}
This section focuses on the precise computation of topological numbers for 2-qubit Hamiltonians \eqref{eq:hp} that exhibit time-reversal symmetry. To achieve this, we introduce an analogous quantum graph model characterized by a tight-binding lattice encompassing four sites. This model facilitates the determination of topological integers \cite{goft2023} and also provides a full description of the related topological phases.

%%%%%%%%%%%%%%%%%%%%%%%%%%%%%%
\subsection{Cartan Hamiltonians and tight-binding quantum graphs \label{sec:quantum_graph}}
The Cartan decomposition of Lie groups \eqref{eq:khaneja_evop} suggests that Hamiltonians $H_p\in i\mathfrak{p}$, which are symmetric under time reversal, can be transformed equivalently using local operations into
\begin{equation}
H_{AI}=\alpha\sigma_{x}\otimes\sigma_{x}+\beta\sigma_{y}\otimes\sigma_{y}+\gamma\sigma_{z}\otimes\sigma_{z} \label{eq:2qHkhaneja}
\end{equation}
where $\alpha,\beta,\gamma\in\mathbb{R}$. There exists a complete topological correlation between $H_{AI}$ and other 2-qubit Hamiltonians within class AI. In other words, the symmetric space for 2-qubits, as expressed in \eqref{eq:R_0 finite}, can be partitioned into subspaces that are equivalent and linked via local transformations, with $H_{AI}$ producing one of these subspaces, as detailed in section \ref{sec:Cartan}.

\begin{figure}
\begin{centering}
\includegraphics[width=0.22\textwidth]{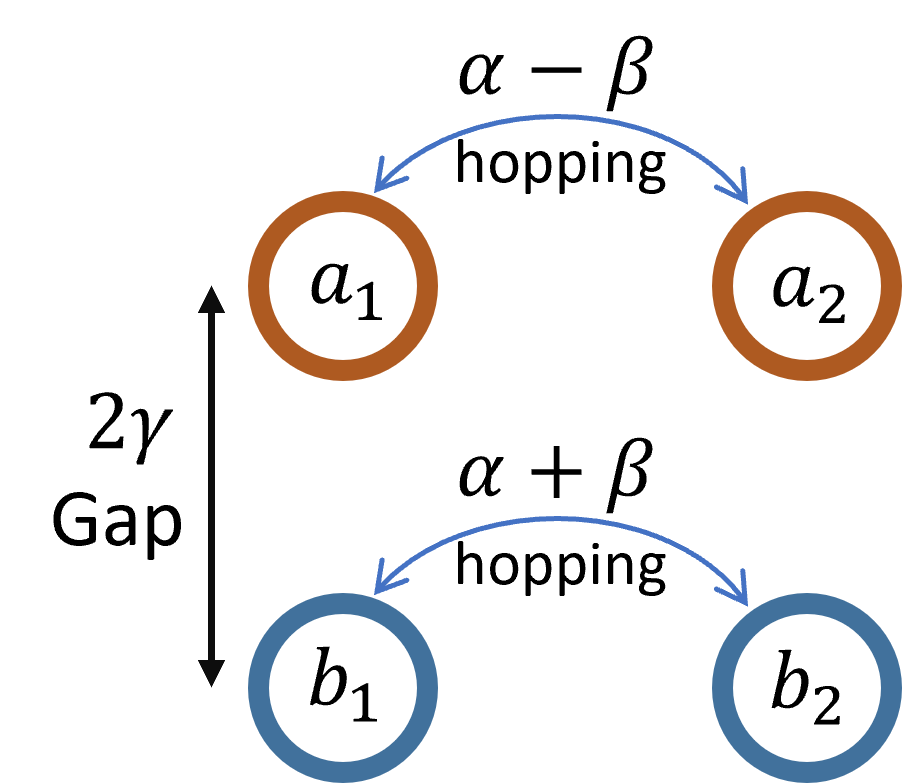}
\par\end{centering}
\caption{ Tight-binding lattice equivalent to the time-reversal symmetric 2-qubit system. This lattice is composed of four sites, organized into two distinct pairs: $a_{1,2}$ and $b_{1,2}$. These pairs are isolated from each other and exhibit an energy difference of $2\gamma$. Within each pair, the two sites are connected by means of a hopping term. \label{fig:4_sites}}
\end{figure}

Using the four states defining the computational basis and the vacuum state $\left|V\right\rangle$, the set of  ladder operators,  
\begin{equation}
  a_{1}^{\dagger}=\left|00\right\rangle \left\langle V\right|,\, a_{2}^{\dagger}=\left|11\right\rangle \left\langle V\right|,\,b_{1}^{\dagger}=\left|01\right\rangle \left\langle V\right|,\, b_{2}^{\dagger}=\left|10\right\rangle \left\langle V\right| 
\end{equation}
allows us to obtain an equivalent description of $H_{AI}$ by means of a  tight-binding model of a particle on a 4-site lattice with the Hamiltonian, 
\begin{equation}
    \begin{split}
H_{\mathrm{graph}}& =\gamma a_{i}^{\dagger}a_{i}+\left(\alpha-\beta\right)\left(a_{2}^{\dagger}a_{1}+a_{1}^{\dagger}a_{2}\right)\\&-\gamma b_{i}^{\dagger}b_{i}+\left(\alpha+\beta\right)\left(b_{2}^{\dagger}b_{1}+b_{1}^{\dagger}b_{2}\right)
\end{split} \label{eq:H_graph}
\end{equation}
which describes two copies of a two-site lattice, each with a different hopping term ($\alpha,\beta$), and a different self-energy ($\gamma$) as displayed in Figure \ref{fig:4_sites}. This mapping between a 4-level quantum system and a lattice is an example of quantum graph  \cite{berkolaiko2013introduction,berkolaiko2017elementary}. Here, the Hamiltonian $H_{AI}$ plays a role similar to an adjacency matrix. $(H_{AI})_{ii}$ is the self-energy of site $i$. $(H_{AI})_{ij} =0 $ for $i\neq j$ that is, for disconnected sites and equals the hopping strength for connected sites. 
The unitary transformation 
\begin{equation}
\left(\begin{array}{c}
a_{j}\\
b_{j}
\end{array}\right)=\frac{1}{\sqrt{2}}\left(\begin{array}{c}
\tilde{a}_{0}+e^{i\pi j}\tilde{a}_{\pi}\\
\tilde{b}_{0}+e^{i\pi j}\tilde{b}_{\pi}
\end{array}\right)=\frac{1}{\sqrt{2}}\sum_{k=0,\pi}e^{ikj}\left(\begin{array}{c}
\tilde{a}_{k}\\
\tilde{b}_{k}
\end{array}\right) \label{eq:Bloch_basis}
\end{equation}
diagonalizes $H_{\mathrm{graph}}$,
 \begin{equation}
     H_{\mathrm{graph}}=\left(\begin{array}{cc}
\tilde{a}_{k} & \tilde{b}_{k}\end{array}\right)
\mathcal{H}\left(k\right)
\left(\begin{array}{c}
\tilde{a}_{k}\\
\tilde{b}_{k}
\end{array}\right)
 \end{equation}
 where 
\begin{equation}
    \mathcal{H}\left(k\right)=\left(\begin{array}{cc}
\gamma+\left(\alpha-\beta\right)\cos k & 0\\
0 & -\gamma+\left(\alpha+\beta\right)\cos k
\end{array}\right). \label{eq:symbol}
\end{equation}
Given that the 4-sites lattice is arranged into two separate pairs, one can decompose $\mathcal{H}\left(k\right)= \mathcal{H}_1 \left(k\right) \oplus \mathcal{H}_2 \left(k\right)$, where $\mathcal{H}_1 \left(k\right) = \gamma + (\alpha - \beta)\cos k$ and $\mathcal{H}_2 \left(k\right) = -\gamma + (\alpha + \beta)\cos k$.
%Note that the unitary transformation \eqref{eq:Bloch_basis} diagonalized $H_{\mathrm{graph}}$. 
%It is identified with the diagonalization proposed above by $n=e^{i\pi j}$ and $m=1\rightarrow a$, $m=2\rightarrow b$. 
Expression (\ref{eq:symbol}) resembles a Bloch Hamiltonian even though this lattice lacks translational symmetry. In fact, $\mathcal{H}\left(k\right)$ is recognized as the Weyl symbol corresponding to the Hamiltonian $H_{\mathrm{graph}}$. The symbol of an elliptic operator is described through its Weyl transform \cite{goft2023}. With translational symmetry, the symbol aligns with the Bloch Hamiltonian, thus explaining the perceived similarity. A notable advantage of the symbol is that it provides a straightforward and systematic approach to determine and compute topological features of a Hamiltonian, as elaborated in \ref{sec:Topological_Invariants}.

Generally, symbols $\mathcal{H}(\boldsymbol{k})$ are represented by $w \times w$ invertible matrices%for $\boldsymbol{k} \neq 0$
. They can be expressed by means of a set of $(p,q+1)$ anti-commuting Dirac matrices $\boldsymbol{\gamma}_{s,a}$ squaring to one \cite{goft2023}, under the form:
\begin{equation}
\label{eq:Diracdefnition}
\mathcal{H}(\boldsymbol{k}) \equiv \boldsymbol{h}_s \cdot \boldsymbol{\gamma}_s + \boldsymbol{h}_a \cdot \boldsymbol{\gamma}_a \, .
\end{equation}
The components $\boldsymbol{h}_{s,a}$ are symmetric and anti-symmetric under $\boldsymbol{k} \to -\boldsymbol{k}$:
$\boldsymbol{h}_s(\boldsymbol{k}) = \boldsymbol{h}_s(-\boldsymbol{k}),\;\;\;
\boldsymbol{h}_a(\boldsymbol{k}) = -\boldsymbol{h}_a(-\boldsymbol{k})$.
The matrices $\boldsymbol{\gamma}_{s,a}$ are generators of a Clifford algebra $Cl_{q+1,p}$, %satisfying:
%$\gamma_i \gamma_j + \gamma_j \gamma_i = \pm \delta_{ij}$. The 
and their size $w $ is
\begin{equation}
\label{eq:clifford_dimension}
w = \begin{cases}
2^{\frac{p+q}{2}}, & p+q \text{ is even}, \\
2^{\frac{p+q+1}{2}}, & p+q \text{ is odd}.
\end{cases}
\end{equation}
The relations,
\begin{equation}
    \label{eq:clifford_correspondence}
     \begin{cases}
      p = d + \frac{s }{2}    \;(\text{mod}\, 4) \\
      q = d - \frac{s }{2}  
    \end{cases}
\end{equation}
allow us to determine the symmetry class $s$ and the spatial dimension $d$ and to relate them under the form of a $(s,d)$ classification presented in Table \ref{tab:10-fold_way}.  The Clifford algebra $Cl_{1,0}$ associated with (\ref{eq:symbol}) is derived by choosing $p=q=0$, which leads to $s = p-q = 0$. This defines the class AI symmetry for a time-reversal invariant two-qubit system with $d=0$, confirming the statement in section \ref{sec:Hamiltonian_symmetric_space} that a two-qubit system has zero spatial dimension. Furthermore, since $w =1$, the expression $\mathcal{H}\left(k\right) = \mathcal{H}_1 \left(k\right) \oplus \mathcal{H}_2 \left(k\right)$ simplifies to an effectively $1 \times 1$ matrix. Finally, the class AI for $d=0$ supports, in the stable limit, a $\mathbb{Z}$ topology.

%This represents the unitary block diagonalization discussed in section \ref{sec:tenfold}. It is important to note that $k$ is dimensionless, given the absence of a natural length scale in the initial system. The symbol $\mathcal{H}\left(k\right)$ is determined by the parameter $k$, and consequently, the base space, analogous to the Brillouin zone in condensed matter physics, simplifies to only two points: $k=0$ and $k=\pi$. This implies that the base space is a 0-dimensional sphere $S^0$, thereby supporting the assertion made in section \ref{sec:Hamiltonian_symmetric_space} regarding the zero-dimensional nature of 2-qubit systems.

%\comment{ more explanations about the symbol and Bloch? }

%%%%%%%%%%%%%%%%%%%%%%%%%%%%%%%%%%%%%%%%%%%%%%%%%%%%%%%%%%%%%%%%%%%%%%%%%%%%%%%%%

\subsection{Topological phases and invariants \label{sec:Topological_Invariants}}
As previously discussed, converting $H_{AI}$ in \eqref{eq:2qHkhaneja} into a quantum graph characterized by the tight-binding Hamiltonian $H_{\mathrm{graph}}$ \eqref{eq:H_graph} enables the application of established methods to compute topological invariants \cite{goft2023}. The index theorems \cite{Atiyah1963,Atiyah1968,Nakahara1990,yankowsky2013,Roe2013} suggest that the symbol \eqref{eq:symbol} shares the same topological properties as the original Hamiltonian, including identical topological invariants. However, apart from this topological equivalence, the symbol $\mathcal{H}\left(k\right)$ and the $H_{AI}$ family of Hamiltonians discussed in section \ref{sec:2qubits} are distinct systems with unique attributes. After applying the unitary Weyl transform \eqref{eq:Bloch_basis}, the symbol $\mathcal{H}\left(k\right)$ is decomposed into two independent $1\times 1$ blocks, represented as $\mathcal{H}\left(k\right)=\mathcal{H}_1 \left(k\right) \oplus \mathcal{H}_2 \left(k\right)$. According to (\ref{eq:Diracdefnition}), each block is characterized by a scalar, even function of $k$, namely $\mathcal{H}_{1,2} \left(k\right) = h_s ^{(1,2)} (k) $.
 %, i.e. a $1\times 1$ matrix as required for class AI 
 \cite{Ertem2017,Pires_Topology_book,goft2023}.
%as the real Clifford algebra with a single positive generator, $Cl_{0,0}$ (as expected for class AI). 

The topological numbers $\nu$ associated to $H_{AI}$ are calculated from $h_s ^{(1,2)} (k)$ in (\ref{eq:Diracdefnition}) %$1 \times 1$ matrices 
 for the Clifford algebra $Cl_{1,0}$ and are given by
\begin{equation}
    \nu_i = \frac{1}{\Omega_{S^0}} \sum_{k\in S^0} \frac{h_s ^{(i)}(k)}{\left|h_s ^{(i)}(k)\right|} \label{eq:nu_0d} \, ,
\end{equation}
where $\Omega_{S^0}=2$ is the area of the zero-dimensional unit sphere. We therefore obtain,
\begin{equation}
    \nu=\nu_{1}+\nu_{2} 
\end{equation}
%where $\nu_0(1)$ and $\nu_0 (2)$ are the topological numbers calculated for each block respectively. 
%\bcomment{added:} The topological numbers for a $d$-dimensional system are given by an integral over the Brillouin zone of a differential form constructed from the symbol. In cases of $\mathbb{Z}$ topology or unstable variants (see \ref{sec:finite}), this integral admits a simple form \cite{goft2023}. For a $1\times 1$ Hamiltonian $h(k)$ with a $S^0$ base space (Brillouin zone) the integral is replaced with a sum resulting in 
Applying \eqref{eq:nu_0d} leads to:
\begin{equation}
\begin{split}
    \nu_{1}&=\frac{1}{2}\sum_{k=0,\pi}\frac{\gamma+\left(\alpha-\beta\right)\cos k}{\left|\gamma+\left(\alpha-\beta\right)\cos k\right|}\\
   & =\begin{cases}
1 & \gamma>\left(\alpha-\beta\right)\\
0 & -\left(\alpha-\beta\right)<\gamma<\left(\alpha-\beta\right)\\
-1 & \gamma<-\left(\alpha-\beta\right)
\end{cases}    
\end{split}
\end{equation}
\begin{equation}
    \nu_{2} =\begin{cases}
1 & \gamma>\left(\alpha+\beta\right)\\
0 & -\left(\alpha+\beta\right)<\gamma<\left(\alpha+\beta\right)\\
-1 & \gamma<-\left(\alpha+\beta\right)
\end{cases}
\end{equation}
such that the set of topological numbers is  
\begin{equation}
    \nu\in\left\{ \pm2,\pm1,0\right\} \simeq\mathbb{Z}_{5}
\end{equation}
as announced in \eqref{eq:pi0R04}  using a different analysis based on the symmetric space  \eqref{eq:R_0 finite} of Hamiltonians. 

With this collection of topological numbers at our disposal, we proceed to examine the potential transitions between the phases they define. Diagonalizing the Hamiltonian (\ref{eq:H_graph}), leads to:
\begin{equation}
    H_{\mathrm{graph}}= \sum_{m=1,2}\sum_{n=\pm}\lambda_{m,n}\left|m,n\right\rangle \left\langle m,n\right|
\end{equation}
where $\left|1,\pm\right\rangle =\frac{1}{\sqrt{2}}\left(\left|00\right\rangle \pm\left|11\right\rangle \right)$, $\left|2,\pm\right\rangle =\frac{1}{\sqrt{2}}\left(\left|01\right\rangle \pm\left|10\right\rangle \right)$ are Bell states and $\lambda_{1,\pm}=\gamma\pm\left(\alpha-\beta\right)$, $\lambda_{2,\pm}=-\gamma\pm\left(\alpha+\beta\right)$.
The transition between distinct topological numbers $\nu$ is characterized by the emergence of a zero-energy mode, specifically identified by the condition where an eigenenergy $\lambda_{m,n}=0$. This phenomenon aligns with index theorems \cite{Atiyah1963, Atiyah1968, Nakahara1990, yankowsky2013,Roe2013,Stone1984}, which assert that each topological number related to the symbol  $\mathcal{H}\left(k\right)$ must correspond to a zero-energy mode $\nu_{ZM}$ of the related Hamiltonian $H_{\mathrm{graph}}$.  

\begin{figure}
\begin{centering}
\includegraphics[width=0.47\textwidth]{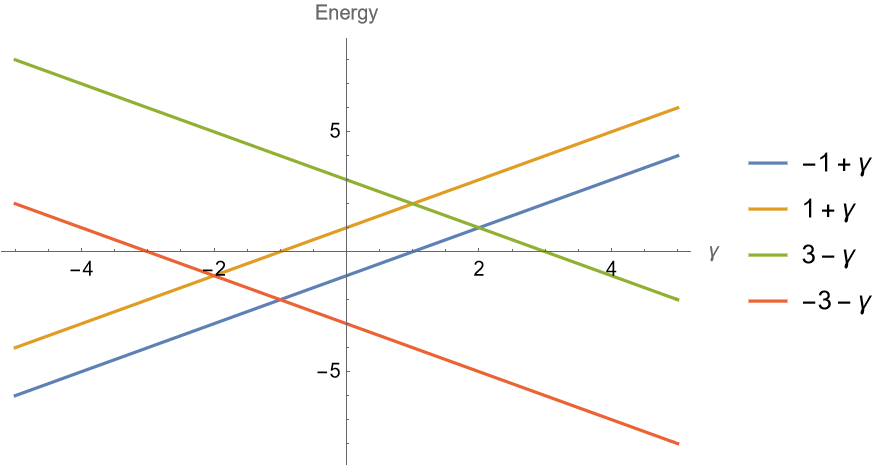}
\par\end{centering}
\caption{The 4 eigenenergies $\lambda_{i,\pm}\left(\gamma\right)$ of $H_{\mathrm{graph}}$ for $\alpha=1$, $\beta=3$. The legend is in the same order as the vectors, i.e. the blue line is for $\left|1,+\right\rangle $, the orange for $\left|1,-\right\rangle $, etc. For $\left|x\right|>3$, we see bands: energies of $\left|1,\pm\right\rangle $ and $\left|2,\pm\right\rangle $ separated. $\gamma=\pm\left(\alpha+\beta\right),\pm\left(\alpha-\beta\right)$ correspond to a vanishing energy. \label{fig:phasetrans}}
\end{figure}

In this context, precision is required. Index theorems establish a relationship between the index of elliptic operators on compact manifolds without boundaries and the topological integers derived from their symbols. Although the index of hermitian elliptic operators like $H_{\mathrm{graph}}$ is inherently zero, $H_{\mathrm{graph}}$ can be expressed using a pair of non-hermitian conjugate operators, $\mathcal{D}$ and $\mathcal{D}^\dagger$, whose index, $\text{Index} \, \mathcal{D} \equiv \dim \ker \mathcal{D} - \dim \ker \mathcal{D}^{\dagger}$, is finite and relevant to the theorem. It can be easily verified that any zero mode of $\mathcal{D}$ or $\mathcal{D}^\dagger$ is also a zero mode of $H_{\mathrm{graph}}$ \cite{Stone1985}, leading to the expression
\begin{equation}
    \nu_{ZM} = \nu = \text{Index} \, \mathcal{D}  \, .
    \label{ASIT}
\end{equation}
At an interface between phases with distinct topological numbers $\nu$, a zero energy mode ($\lambda_{m,n}= 0$) emerges.  Furthermore, the changes in $\nu_{1}$ and $\nu_{2}$ correspond respectively to $\lambda_{1,\pm}=0$ and  $\lambda_{2,\pm}=0$. In other words, the zero-energy state belongs to the subsystem that changes its topological number, namely a change in $\nu_1$ corresponds to one of the states $\left|1,\pm\right\rangle$ as the zero-energy mode as displayed in figure \ref{fig:phasetrans}. 

Those zero-energy states are the zero-dimensional analogs of topological edge states. It is known \cite{Stone1984,Stone1985,Kitaev2009} that when two
systems with different topological numbers share a common boundary, in-gap zero-energy states show up that are spatially localized on the boundary. %For a zero-dimensional system this translates as changing the system's parameters such that it shifts between different topological numbers in accordance with our results. This is in agreement with the appearance of zero-energy states at the boundaries resulting from the Atiyah-Singer index theorem \comment{citation}, which also \bcomment{involves} 0-dimension base spaces. 

%%%%%%%%%%%%%%%\
%%%%%%%%%%%%%%%%%%%%%%%%%%%%%%%%%%%%%%%%%%%%%%%%%
%%%%%%%%%%%%%%%%%%%%%%%%%%%%%%%%%%%%%%%%%%%%%%%%%
%%%%%%%%%%%%%%%
%%%%%%%%%%%%%%%%%%%%%%%%%%%%%%%%%%%%%%%%%%%%%%%%%%%
\section{Applications to quantum computing and condensed matter \label{sec:generalizations}}

%%%%%%%%%%%%%%%%%%%%%%%%%%%%%%%%%%%%%%%%%%%%%%%%%%%

\subsection{Topological considerations in quantum computation \label{sec:QI}}

In earlier sections, our discussion centered on two-qubit systems, although the approaches outlined in sections \ref{sec:Cartan} and \ref{sec:Topology} have broader applicability. The Wootters concurrence that we employ is akin to other entanglement metrics, like entanglement entropy \cite{Wootters1998} or the maximum violation of Bell inequalities \cite{Verstraete2002}. When considering larger groups of qubits \cite{KHANEJA200111,Bullock2004}, the two-qubit concurrence is substituted with various Uhlmann concurrences \cite{Uhlmann2000}, each generating a distinct Cartan decomposition \eqref{eq:Cartan_Decomp}. Given that a quantum system can exhibit up to two different antiunitary symmetries \cite{Moore}, these concurrences are interconnected via unitary transformations or pertain to different symmetry subsystems.
The main concept emphasized is that topology, although a fundamental feature of quantum physics not limited to condensed matter systems, has been overlooked in quantum computing. Consequently, we anticipate that topological classes and invariants will emerge in distinguishable qubits-based quantum algorithms, regardless of how they are implemented.

Quantum dots provide a good platform to implement qubits and to study their entanglement \cite{Vandersypen2007}. The applicability of our results to quantum dots requires validating some assumptions, e.g. distinguishability, spatial dimension, and symmetry class.

An inherent characteristic of quantum dots and condensed matter systems is the presence of spatially localised collective excitations, such as charge or spin densities, which render them suitable candidates for distinguishable qubits. Consequently, the interaction among a finite array of localized quantum dots is essentially zero-dimensional, irrespective of the substrate's dimension. %From a formal viewpoint, the dimension of the system can be defined as the base space of the symbol, as presented in \ref{sec:topo_phases}. 
%The symbol of finite qubit systems, following the construction of a quantum graph similar to \ref{sec:quantum_graph}, will be zero-dimensional, as it is a discrete system.

The quest for symmetry is more intricate. A clear physical illustration is provided by spin-orbit coupling. For a single spin-1/2 excitation in a two-dot (two orbits) system, the effective time reversal symmetry is 
\begin{equation}
    \Theta=(\mathbbm{1}_2 \otimes i\sigma_y)K,\quad \Theta^{2}=-\mathbbm{1}_4 \label{eq:KM_theta}
\end{equation}
where $\mathbbm{1}_2$ pertains to the pseudo-spin orbital degree of freedom and $i\sigma_y$ to the spin part, hence assigning it to the symmetry class AII. The operator $\Theta$ allows for a Cartan decomposition \eqref{eq:antiunitary_cartan}, but with $    \left\langle \psi\left|\Theta\right|\psi\right\rangle \equiv0
$ for any $\left | \psi\right\rangle \in \mathbb{C}^2\otimes \mathbb{C}^2$. This means that this system lacks inherent entanglement. 

This is a general and known property (see e.g. \cite{Messiah} p. 672): Consider an antiunitary operator $\Theta$ that satisfies $\Theta^{2}=-\mathbbm{1}$. It is antihermitian: $\Theta^{\dagger}=\Theta^{-1}=-\Theta$.
From the definition of hermitian conjugation of antilinear operators:
\begin{equation}
\left\langle \psi\right|\left(\Theta\left|\psi\right\rangle \right)=\left\langle \psi\right|\left(\Theta^{\dagger}\left|\psi\right\rangle \right)=-\left\langle \psi\right|\left(\Theta\left|\psi\right\rangle \right),
\end{equation}
implying $\left\langle \psi\left|\Theta\right|\psi\right\rangle \equiv0$ for any state $\ket\psi$. 
From a broader perspective, the entanglement resulting from spin-orbit interactions involves various degrees of freedom within the same excitation or particle (intraparticle entanglement). While measuring this type of entanglement has been suggested \cite{Dasenbrook2016}, conducting a Bell experiment to test it is not feasible \cite{Paneru2020}, because the excitation acts as a single particle. This situation can be improved by including more particles, such as physical qubits. For instance, by introducing an additional particle, the time-reversal symmetry is expressed as
\begin{equation}
    \Theta_{QD}=(\mathbbm{1}_2 \otimes i\sigma_y \otimes i\sigma_y)K.
\end{equation}
As a result, $\Theta_{QD}^2=+\mathbbm{1}_8$. The system falls into class AI, resembling a two-qubit system, and allows for entanglement between particles  or their spins. The procedure of merging two systems, each having a symplectic antiunitary symmetry $(\Theta^2=-\mathbbm{1})$, to form a larger one with $ \Theta^2=+\mathbbm{1}$ is a general strategy \cite{D'Alessandro}. This technique also provides insight into the concept of multiple concurrences for a single system \cite{Uhlmann2000}, which defines the entanglement of subsystems. Examples involving quantum dots demonstrate that the connection between topology and entanglement persists beyond purely theoretical qubit systems. 

%%%%%%%%%%%%%%%%%%%%%%%%%%%%%%%%%%%%%%%%%%%%%%%%%
%%%%%%%%%%%%%%%
%%%%%%%%%%%%%%%%%%%%%%%%%%%%%%%%%%%%%%%%%%%%%%%%%%%
\subsection{Qubits in condensed matter \label{sec:condmat}}

We proceed to examine the implications of our findings concerning quantum entanglement, as characterized by the Wootters concurrence (\ref{eq:concurrence}), within the realm of condensed matter systems. In the study of entanglement within condensed matter physics, we aim to identify effective two-qubit systems \cite{Chtchelkatchev2002,Beenakker2003,Costa2006,Vandersypen2007,doucot2008,Samuelsson2009,Roche2020,Beenakker2006}, striving to simplify a complex and large many-particle Hilbert space to the dimensions of the two-qubit $\mathbb{C}^2\otimes \mathbb{C}^2$ Hilbert space. Quantum dots serve as examples of qubit counterparts that are isolated from the continuous spectrum characteristic of the underlying condensed matter system. Two primary types of analogous qubits are recognized. The first type involves systems with a few discrete states, such as seen in quantum dots \cite{Beenakker2003,Costa2006}, separated from the continuous Hilbert space. In the second type, the Hilbert space results from the continuous multiplication of an infinite series of finite spaces \cite{Chtchelkatchev2002,Hannes2008,doucot2008,Roche2020}, with 2-qubits making up typical excitations.

In the initial scenario, indistinguishable fermions fill the discrete states, turning the problem into one of distinguishable particles by tracing out the continuous host. Quantum entanglement between different particles emerges. Topological insulators, which support discrete and protected states, naturally suit the implementation of qubit physics.

In the latter scenario, the repeated multiplication of non-interacting subspaces emerges due to translational symmetry and Bloch's theorem. This indicates that any spatial disturbance, such as a defect, will connect these subspaces, thereby disrupting any qubit description. A brief calculation illustrating this entanglement loss is detailed in appendix \ref{app:Entanglement_Decay}. Furthermore, it is essential to highlight that, generally, the entanglement discussed involves the internal degrees of freedom within a single excitation (intraparticle). Describing these correlations as quantum entanglement is inadequate (section \ref{sec:QI}).

To apply previous results in the realm of condensed matter systems, it is essential to consider two additional aspects: spatial dimensions and indistinguishable fermions. Topological invariants have been calculated and assessed for 2-qubit configurations in zero-dimensional systems, as elaborated in section \ref{sec:2qubits}. The findings shown in section \ref{sec:2qubits} for the 2-qubit system's topological invariant rely on this zero-dimensional nature. As highlighted in section \ref{sec:Hamiltonian_symmetric_space}, the tenfold classification has been adapted to incorporate the spatial dimension pertinent to condensed matter systems. This adaptation is illustrated for dimensions $d=0,1,2,3$ in table \ref{tab:10-fold_way}, and can be generalized for any dimension using relation \eqref{eq:pii_Ri}.  

\begin{table}
% \small
\begin{center}\begin{tabular}{|S|S|S|S|S|S|S|S|S|}
\hline 
\text{Class} & $\Theta$ & $\Pi$ & $S$ & $d=0$ & 1 & 2 & 3
\\
\hline 
\hline 
A  & 0 & 0 & 0 & \cellcolor{cyan!10}$\mathbb{Z}$ & 0 & \cellcolor{cyan!10}$\mathbb{Z}$ & 0
\\
\hline 
AIII  & 0 & 0 & 1 & 0 & \cellcolor{cyan!10}$\mathbb{Z}$ & 0 & \cellcolor{cyan!10}$\mathbb{Z}$
\\
\hline 
\hline
AI  & + & 0 & 0 & \cellcolor{cyan!10}$\mathbb{Z}$ & 0 & 0 & 0
\\
\hline 
BDI  & + & + & 1 & \cellcolor{cyan!10}$\mathbb{Z}_2$ & \cellcolor{cyan!10}$\mathbb{Z}$ & 0 & 0
\\
\hline 
D  & 0 & + & 0 & \cellcolor{cyan!10}$\mathbb{Z}_2$ & \cellcolor{cyan!10}$\mathbb{Z}_2$ & \cellcolor{cyan!10}$\mathbb{Z}$ & 0
\\
\hline 
DIII  & \textbf{--} & + & 1 & 0 & \cellcolor{cyan!10}$\mathbb{Z}_2$ & \cellcolor{cyan!10}$\mathbb{Z}_2$ & \cellcolor{cyan!10}$\mathbb{Z}$
\\
\hline 
AII  & \textbf{--} & 0 & 0 & \cellcolor{cyan!10}$2\mathbb{Z}$ & 0 & \cellcolor{cyan!10}$\mathbb{Z}_2$ & \cellcolor{cyan!10}$\mathbb{Z}_2$
\\
\hline 
CII  & \textbf{--} & \textbf{--} & 1 & 0 & \cellcolor{cyan!10}$2\mathbb{Z}$ & 0 & \cellcolor{cyan!10}$\mathbb{Z}_2$
\\
\hline 
C  & 0 & \textbf{--} & 0 & 0 & 0 & \cellcolor{cyan!10}$2\mathbb{Z}$ & 0
\\
\hline 
CI  & + & \textbf{--} & 1 & 0 & 0 & 0 & \cellcolor{cyan!10}$2\mathbb{Z}$
\\
\hline
\end{tabular}
\end{center}
\caption{Tenfold classification. The first four columns display the 10 symmetry classes defined by their antiunitary symmetries $\Theta,\,\Pi$ and chirality $S$, the same as in \ref{table:10fold_with_Ri}. The last 4 columns indicate the relevant homotopy group ($0$, $\mathbb{Z}$, or $\mathbb{Z}_2$) of the corresponding stable symmetric space (see section \ref{sec:Topology}) as a function of the dimension $d$. As expected, the $d=0$ column coincides with the last column of table \ref{table:10fold_with_Ri}. \label{tab:10-fold_way}}
\end{table}

Since any weakly-interacting system of fermions is described by the tenfold classification (see section \ref{sec:tenfold}), it involves natural antiunitary symmetries. We anticipate that using these symmetries to define concurrence and perform a Cartan decomposition will lead to results similar to section \ref{sec:Cartan}. 

It is important to highlight that not all antiunitary operators proposed by the tenfold way qualify as candidates for concurrence. Consider, for instance, the Kane-Mele model \cite{Kanemele2005}, characterized by $d=2$ and belonging to class AII in the tenfold classification \ref{tab:10-fold_way}. This model respects only time reversal symmetry \eqref{eq:KM_theta}, where the orbital degrees of freedom represent the graphene pseudo-spin. As noted earlier, this operator does not allow defining concurrence.
By utilizing natural symmetries to define concurrence, we ensure that Hamiltonians respecting these symmetries are entangling operators. When considering the tenfold classification (table \ref{table:10fold_with_Ri} or \ref{tab:10-fold_way}), this implies that systems in the AII, CII, or C classes do not possess a clear definition of concurrence. In other words, a system could be topologically non-trivial and described through the tenfold way but still lack entanglement possibilities. Conversely, if concurrence is defined through an external operator, the resulting Hamiltonians may include both entangling and non-entangling operators.

In condensed matter physics, we deal with indistinguishable fermions. Nonetheless, a 2-qubit framework inherently refers to distinguishable particles. Considering quantum dots as an example, allowing two fermion excitations within a single quantum dot, their fermionic characteristics become prominent, thus losing the 2-qubit correlation. Specifically, 2 fermions occupying 2 quantum dots, as cited in \cite{Schliemann2001,Eckert2002}, cannot be simplified into a 2-qubit model. Such a scenario involving 2 fermions in 2 quantum dots naturally exhibits particle-hole symmetry, and consequently, concurrence based on this symmetry has been suggested:
\begin{equation}
    C_{PH}\equiv\left|\bra{\psi}\Pi\ket{\psi}\right|, \label{eq:conc_ph}
\end{equation} 
where $\Pi$ is the particle-hole operator. In the absence of time reversal symmetry, similar arguments to those in section \ref{sec:Cartan} demonstrate that this system's entangling operators (as defined in section \ref{sec:Cartan}) belong to Cartan class D for $d=0$. Notably, this indicates the system is topologically non-trivial, and it remains so even if time reversal is restored, transitioning to class BDI (see table \ref{table:10fold_with_Ri} for both). Based on these ideas, we may conclude that the Cartan decomposition and tenfold classification remain applicable frameworks for any fermionic system.

%In section \ref{sec:quantum_graph} we have shown there is an equivalence between the 2-qubit system and fermions on a 4-site lattice. There, the fermionic nature of the particles has no effect on the physical properties. 

These concepts render topological modes in condensed matter an attractive option for qubits with topological protection, particularly those induced by defects \cite{Teo2010,goft2023}, due to their ease of manipulation. Fermions present in these modes can interact through Coulomb forces, leading to a topological version of the Heisenberg model.

%%%%%%%%%%%%%%%%%%%%%%%%%%%%%%%%%%%%%%%%%%%%%%%%%%%%
%%%%%%%%%%%%%%%%%%%%%%%%%%%%%%%%%%%%%%%%%%%%%%%%%%%
%%%%%%%%%%%%%%%%%%%%%%%%%%%%%%%%%%%%%%%%%%%%%%%%%%%
\section{Conclusions}

The purpose of this paper was to link quantum entanglement with topology, particularly demonstrating that for two distinguishable qubits in zero spatial dimension, this connection arises from anti-unitary symmetries. Quantum entanglement, quantified by Wootters concurrence or similar methods, depends on anti-unitary symmetries like time reversal or particle-hole. These symmetries enable the Cartan decomposition of 2-qubit operators into entangling and non-entangling gates and provide a topological classification for Hamiltonians and time evolution operators in large Hilbert spaces. Since 2-qubit systems do not always belong to that stable limit, we have extended these topological considerations to finite and low-dimensional Hilbert spaces along two complementary approaches. One based on the identification of homotopy groups and the second using the symbol, a quantity playing a central role in the family of index theorems. We have devised a method for computing topological numbers, identifying topological phases, and finding protected zero modes by mapping 2-qubit Hamiltonians to quantum graphs.

A surprising finding is the strong link between entanglement and topology. For entangling quantum gates, it is essential, but not enough, for the symmetric space of a Hamiltonian family to have non-trivial topology; Alice and Bob must be topological.

%We have shown for the 2-qubit that the Cartan decomposition based on an antiunitary symmetry defines entangling operators based on a given definition of concurrence. This Cartan decomposition also determines the topology, thereby connecting it to the entanglement.

%We have shown that the 2-qubit system is topologically non-trivial regardless of implementation, i.e. Alice and Bob are topological. This topology is revealed at the level of the Hamiltonian via non-zero topological numbers, accounting for different topological phases. These topological invariants are described by the Cartan tenfold classification. 

The finiteness of the 2-qubits Hilbert space allows for a richer variety of topological numbers as compared to the tenfold classification established in the stable limit of very large systems.%, and also allows prediction of the number of physically available topological phases. These are computed by a careful approach consisting of identifying the symmetric spaces from physical constraints. We have devised a method of mapping to quantum graphs, allowing the systematic computation of topological numbers and phase transitions. 

For time evolution operators, i.e. quantum gates, a relation has been obtained between topological numbers and a geometric (Berry) phase. Thus, beyond its ensuring entanglement, topology allows also to measure concurrence using the relation between geometric phase and concurrence.

These concepts may be broadened to include fewer constraints, such as higher spatial dimensions or various symmetry classes, yielding a tenfold table \ref{tab:10-fold_way} that reveals topological characteristics beyond stable symmetric spaces.

Studying entanglement in many-body condensed matter physics often involves using qubit analogue systems. In contrast, we propose leveraging topological modes in condensed matter systems to develop protected, distinguishable qubits. A prime example is the Hubbard model, where Coulomb interaction-induced strong correlations enable electronic entanglement. % in molecules (\comment{citation? Amit's thesis?}). Topologically protected discrete modes can play the role of such orbitals. One can now search for a protocol where these effective condensed matter qubits can be entangled.

%Further questions regarding the efficiency, tolerance and error corrections in gates constructed by topological materials are yet to be answered.

%%%%%%%%%%%%%%%
%%%%%%%%%%%%%%%%%%%%%%%%%%%%%%%%%%%%%%%%%%%%%%%%%%%

\begin{acknowledgments}
We thank B. Rotstein for insightful discussions and C. L. Schochet for his assistance with the understanding  of symmetric spaces. We also thank Y. Abulafia for her help in defining zero modes, and A. Goft for lending us symbols. This research was funded by the Israel Science Foundation Grant No.~772/21 and the Pazy Foundation. N.O. acknowledges the support received through the VATAT Scholarship Program for PhD Students in Quantum Science and Technology.
\end{acknowledgments}

%%%%%%%%%%%%%%%%%%%%%%%%%%%%%%%%%%%%%%%%%%%%%%%%%%%%%
%%%%%%%%%%%%%%%%%%%%%%%%%%%%%%%%
\appendix

\section{Cartan Decomposition of Evolution Operators \label{app:khaneja}}
We present a proof for the Cartan decomposition of evolution operators \eqref{eq:khaneja_evop} discussed in section \ref{sec:Cartan}. 

We first show (Cartan decomposition of evolution operators) that for an operator $\mathcal{O}\in U(4)$ there exists $\left(k_0,\vec{k}\right)\in \mathbb{R}^4$ such that
\begin{equation}
    \mathcal{O} =\left(A_{1}\otimes A_{0}\right)e^{i\left(k_{0}\mathbbm{1}+\vec{k}\cdot\vec{\Sigma}\right)}\left(B_{1}\otimes B_{0}\right) \label{eq_app:khaneja}
\end{equation}
where $A_i,B_i\in SU(2)$ and $\vec\Sigma_i=\sigma_i\otimes\sigma_i$.

We will use (without proof) three known results:
\begin{enumerate}
\item R1: There exists a mapping $\Phi:SU\left(2\right)\times SU\left(2\right)\rightarrow SO\left(4\right)$:
\[
\left(A,B\right)\rightarrow\Phi\left(A,B\right)=M^{\dagger}\left(A\otimes B^{*}\right)M,
\]
where $M$ is the unitary Makhlin matrix \cite{Makhlin2000}:
\[
M=\frac{1}{\sqrt{2}}\left(\begin{array}{cccc}
1 & 0 & 0 & i\\
0 & i & 1 & 0\\
0 & i & -1 & 0\\
1 & 0 & 0 & -i
\end{array}\right). \label{eq_app:magic_makhlin}
\]
which is also the basis transfer matrix defining the Wootters magic basis \cite{Hill1997}. $\Phi$ is a 2 to 1 homomorphism, since $\Phi\left(A,B\right)=\Phi\left(-A,-B\right)$. It is onto, meaning there is an inverse map to a selection of phase.
\item R2: Let $\mathcal{O}$ be a unitary matrix. Define $\mathcal{O}_{R}=\left(\mathcal{O}+\mathcal{O}^{*}\right)/2$,
$\mathcal{O}_{I}=\left(\mathcal{O}-\mathcal{O}^{*}\right)/2$. Then $\mathcal{O}_{R}$ and $\mathcal{O}_{I}$ are real
matrices and
\[
Q=\left(\begin{array}{cc}
\mathcal{O}_{R} & \mathcal{O}_{I}\\
-\mathcal{O}_{I} & \mathcal{O}_{R}
\end{array}\right)
\]
is orthogonal. Moreover, $\mathcal{O}_{R}\mathcal{O}_{I}^{T}$ and $\mathcal{O}_{R}^{T}\mathcal{O}_{I}$ are
real symmetric.
\item R3: Let $A,B$ be two real rectangular matrices. Then
there exist 2 orthogonal matrices $U,V$ such that:
\[
D_{1}=U^{T}AV,\quad D_{2}=U^{T}BV
\]
are real diagonal matrices iff $AB^{T}$ and $A^{T}B$ are symmetric.
\end{enumerate}

Corollary: for a unitary matrix $\mathcal{O}$, there exist orthogonal matrices $Q_L$ and $Q_R$ and a diagonal unitary matrix $D_\mathcal{O}=\mathrm{diag}\left(e^{i\theta_1},e^{i\theta_2},e^{i\theta_3},e^{i\theta_4}\right)$ such that 
\begin{equation}
    \mathcal{O}=Q_L D_\mathcal{O} Q_R.
\end{equation}
Proof: Define $\mathcal{O}_R$ and $\mathcal{O}_L$ as in R2. Then from R2, $AB^{T}$ and $A^{T}B$ are real symmetric. By R3 there exist orthogonal matrices $Q_L$ and $Q_R$ such that 
\begin{equation}
    D_{R}=Q_{L}^{T}\mathcal{O}_{R}Q_{R},\quad D_{I}=Q_{L}^{T}\mathcal{O}_{I}Q_{R}
\end{equation}
where $D_{R}$ and $D_{L}$ are real diagonal matrices. Thus $D_\mathcal{O}=Q_{L}^{T}\mathcal{O}Q_{R}$ is diagonal since $\mathcal{O}=\mathcal{O}_R+i\mathcal{O}_I$. Since $\mathcal{O}$ is unitary, $D_\mathcal{O}=D_R+iD_I$ is unitary and diagonal. 

Now we are in a position to establish \eqref{eq_app:khaneja}: Let $\mathcal{O}^\prime = M \mathcal{O} M^\dagger$ with $M$ defined in R1. $\mathcal{O}^\prime$ is unitary and according to the corollary there exist orthogonal matrices $Q_L$ and $Q_R$ and a diagonal unitary matrix $D_\mathcal{O}$ such that
\begin{equation}
    \mathcal{O}^\prime =Q_L D_\mathcal{O} Q_R^T.
\end{equation}
with $D_\mathcal{O}=\mathrm{diag}\left(e^{i\theta_1},e^{i\theta_2},e^{i\theta_3},e^{i\theta_4}\right)$. 

According to R1, there are $A_1,A_2,B_1,B_2\in SU(2)$ such that
\begin{equation}
    \begin{split}
MQ_{L}M^{\dagger}=A_{1}\otimes A_{0}\\
MQ_{R}M^{\dagger}=B_{1}\otimes B_{0}
\end{split}
\end{equation}
such that
\begin{equation}
    \mathcal{O} = \left(A_{1}\otimes A_{0}\right)M D_\mathcal{O} M^\dagger\left(B_{1}\otimes B_{0}\right).
\end{equation}
By a direct calculation:
\begin{equation}
    M D_\mathcal{O} M^\dagger=e^{i\left(k_{0}\mathbbm{1}+\vec{k}\cdot\vec{\Sigma}\right)}
\end{equation}
where  $\vec\Sigma_i=\sigma_i\otimes\sigma_i$ and $k_i$ are related to $\theta_i$ by:
\begin{equation}
    \left(\begin{array}{c}
\theta_{0}\\
\theta_{1}\\
\theta_{2}\\
\theta_{3}
\end{array}\right)=\left(\begin{array}{cccc}
1 & 1 & -1 & 1\\
1 & 1 & 1 & -1\\
1 & -1 & -1 & -1\\
1 & -1 & 1 & 1
\end{array}\right)\left(\begin{array}{c}
k_{0}\\
k_{1}\\
k_{2}\\
k_{3}
\end{array}\right).
\end{equation}
thus $\left(k_0,\vec{k}\right)\in \mathbb{R}^4$, which concludes the proof.

%At first glance, this proof appears unrelated to the Cartan decomposition. However, this is just superficial. In the magic basis, defined by \eqref{eq_app:magic_makhlin}, the antinunitary operator is $\Theta=K$, as referenced in \ref{sec:2qubits}. Within this basis, the decomposition. 

%%%%%%%%%%%%%%%%%%%%%%%%%%%%%%%%%%%%%%%%%%%%%%%%%%%%%
%%%%%%%%%%%%%%%%%%%%%%%%%%%%%%%%

\section{Breaking a 2-qubit analogue in continuous Hilbert spaces \label{app:Entanglement_Decay}}
In this appendix, we consider a specific example of defect so as to illustrate our claim. To that purpose, consider graphene, a two-dimensional material in the presence of spin-orbit coupling (SOC) as defined by a Rashba term. This example is sometimes presented as a 2-qubit analogue \cite{Roche2020}. We show that 2-qubit-like excitations decay exponentially fast with time when a single-site localized perturbation is added.

The low energy and continuous limit Bloch Hamiltonian for a single valley in graphene with Rashba SOC is \cite{Roche2020}:
\begin{equation}
    \mathcal{H}_0 (\boldsymbol{k})=\hbar v_F \left(\sigma_x k_x +\sigma_y k_y\right)\otimes I + \lambda_R\left( \sigma_x\otimes\sigma_y +\sigma_y\otimes\sigma_x\right)
\end{equation}
Now, consider a localized perturbation represented by an adatom \cite{Dutreix2019}, where a foreign atom bonds to a carbon atom within the graphene structure. This perturbation is modeled by a Hamiltonian with the adatom assimilated to a Gaussian function of width $\sigma\sim 10^{-10}m$ (indicative of bond length).:
\begin{equation}
    H=
    \sum_{\boldsymbol{k}} c_{\boldsymbol{k}}^{\dagger}
    \mathcal{H}_0 (\boldsymbol{k}) c_{\boldsymbol{k}}
+\sum_{\boldsymbol{k},\boldsymbol{k}^{\prime}} V_0 e^{2\sigma^2 (\boldsymbol{k}-\boldsymbol{k}^{\prime})^2}a_{\boldsymbol{k}^{\prime}}^{\dagger}a_{\boldsymbol{k}} \label{eq_app:graphene_adatom}
\end{equation}
where $c_{\boldsymbol{k}}=\left(\begin{array}{cccc}
a_{\boldsymbol{k}\uparrow} & a_{\boldsymbol{k}\downarrow} & b_{\boldsymbol{k}\uparrow} & b_{\boldsymbol{k}\downarrow}\end{array}\right)^{T}$, and $V_0\sim 10eV$ \cite{Dutreix2019}. 

A fully entangled 2-qubit like state \cite{Roche2020} describes the coupling between the sublattice and spin degrees of freedom for a single state  $\vec{k}$ in the Brillouin zone:
\begin{equation}
   \left|\psi_B(\boldsymbol{k})\right\rangle  =\left(a_{\boldsymbol{k}\uparrow}^{\dagger}+b_{\boldsymbol{k}\downarrow} ^{\dagger}\right)\left|0\right\rangle ="\left|00\right\rangle _{\boldsymbol{k}}+\left|11\right\rangle _{\boldsymbol{k}}".
\end{equation}
The Hamiltonian \eqref{eq_app:graphene_adatom} couples this state to other points $\boldsymbol{k}^\prime$ in the Brillouin zone, which is continuous for large systems. The Fermi golden rule:
\begin{equation}     \Gamma_{\boldsymbol{k}\rightarrow\boldsymbol{k}'} = \frac{2\pi}{\hbar}\rho\left(E_{f}\right) V_0 e^{2\sigma^2 (\boldsymbol{k}-\boldsymbol{k}^{\prime})^2}.
\end{equation}
shows that the interaction does not couple spin-like degrees of freedom, so $\rho\left(\boldsymbol{k}\right)=\pi A \left|\boldsymbol{k}\right|$, where $A$ is the area of the graphene sheet. The total transition rate from the initial $\boldsymbol{k}$ rewrites,
\begin{equation}
    \Gamma_{\boldsymbol{k}\rightarrow\text{any}} = \frac{2\pi^2 A}{\hbar}\left|\boldsymbol{k}\right| \int V_0 e^{2\sigma^2 (\boldsymbol{k}-\boldsymbol{k}^{\prime})^2}d\boldsymbol{k}^{\prime} .
\end{equation}
The Fermi energy is of the order of the backgate voltage $E_f\approx eV_b\sim 1eV$ \cite{Dutreix2019}. Since $E_f=\hbar v_f |k|$, $v_f\sim 10^6 m/s$:
\begin{equation}
    \Gamma_{\boldsymbol{k}\rightarrow\text{any}} = 4\pi^2\frac{ A eV_b V_0}{\sigma v_f \hbar^2}   \int e^{2(\boldsymbol{q}-\boldsymbol{q}^{\prime})^2}d\boldsymbol{q}^{\prime} 
\end{equation}
where $\frac{ eV_b V_0}{\sigma v_f \hbar^2}\sim 10^{35} Hz/m^2$, for graphene sheets of size $A\sim 1 \mu m ^2$, the decay time from the target state is approximately $10^{-23} s$. This duration stays short even with smaller perturbation potentials, such as $V_0\sim 10^{-3} eV$, resulting in a decay time near $10^{-19} s$. Typically, adatoms are modelled with $\sigma\rightarrow0$ (a delta function), leading to a linear decrease in decay time as $\sigma\rightarrow0$.

We conclude that any Bell state exponentially loses coherence due to perturbations and disorder. This is in contrast to topological modes, which are insensitive to disorder.

%%%%%%%%%%%%%%%%%%%%%%%%%%%%%%%%%%%%%%%%%%%%%%%%%%%%%
%%%%%%%%%%%%%%%%%%%%%%%%%%%%%%%%%%%%%%%%%%%%%%%%%%%%%%%%%%%%%%%%%%%%%%%%%%%%%%%%%%%%%
%%%%%%%%%%%%%%%%%%%%%%%%%%%%%%%%

\bibliography{Bibliography/proposal}

\end{document}